\documentclass[jkps, preprint, fleqn,showkeys,showpacs,a4paper]{revtex4}
\usepackage{amsmath} 
\usepackage{float}
\usepackage{graphicx ,xcolor}
\usepackage{multirow,newlfont,enumerate} 
\usepackage{epsfig,epstopdf,ifpdf}
\usepackage{latexsym,slashed}
\usepackage{subfigure, float, epsfig}
\usepackage[final]{pdfpages} 
\usepackage{amsmath,amssymb,amsfonts,amsthm}
\usepackage{todonotes} 
\usepackage{tensor} 
\usepackage{titlesec} 
\usepackage{mathtools, array} 
\usepackage{todonotes} 
\usepackage{tensor} 
\usepackage{mathtools, array} 
\usepackage{mathtools, array} 
\theoremstyle{plain}

\newtheorem{prop}{Proposition} 
\theoremstyle{definition} 

\newcommand{\beq}{\begin{equation}}
\newcommand{\eeq}{\end{equation}}
\newcommand{\beqr}{\begin{eqnarray}}
\newcommand{\eeqr}{\end{eqnarray}}

\def\wt{\widetilde}     
\def\lsim{\raise0.3ex\hbox{$\;<$\kern-0.75em\raise-1.1ex\hbox{$\sim\;$}}}
\def\gsim{\raise0.3ex\hbox{$\;>$\kern-0.75em\raise-1.1ex\hbox{$\sim\;$}}}
\def\para{\vspace{0.3cm}\noindent} 
\def\noi{\noindent}
\def\m{\,{\rm m}}
\def\s{\,{\rm s}}  
\def\g{\,{\rm g}}

\begin{document}

\def\wt{\widetilde}     
\def\lsim{\raise0.3ex\hbox{$\;<$\kern-0.75em\raise-1.1ex\hbox{$\sim\;$}}}
\def\gsim{\raise0.3ex\hbox{$\;>$\kern-0.75em\raise-1.1ex\hbox{$\sim\;$}}}
\def\para{\vspace{0.3cm}\noindent} 
\def\noi{\noindent}
\def\m{\,{\rm m}}
\def\s{\,{\rm s}}  
\def\g{\,{\rm g}}

\title{Correlated  Dynamics of Immune Network and  $ sl( 3, \mathbb{R} )$ Symmetry Algebra }

\author{ R. Dutta} 
\affiliation{Department of Physics, Material Science \& Astronomy, Missouri State University } 
\email{Corresponding Author :  rd5e@missouristate.edu / dutta.22@osu.edu} 

\author { A. Stan} 
\affiliation{Department of Mathematics, The Ohio State University  } 
\email{ stan.7@osu.edu}

\begin{abstract}
We observed existence of periodic orbit in immune network under transitive solvable lie algebra.
In this paper, we  focus  to develop condition of  maximal Lie algebra for immune network model  and use that condition to construct vector field of symmetry to study
non linear pathogen model.
We used two methods to obtain analytical structure of solution, namely  normal generator and differential
 invariant function. 
Numerical simulation of analytical structure exhibits correlated periodic pattern growth under spatio temporal symmetry which is similar to linear dynamical simulation result.
We used  lie algebraric method to understand  correlation between growth pattern and  symmetry of dynamical system.
We employ idea of using one parameter point group of transformation  of variables under which  linear manifold is retained.  
In procedure, we present the method of deriving Lie point symmetries, calculation of first integral and invariant solution for the ODE. 
We show the connection between symmetries and differential  invariant solution of the ODE.  
The analytical structure of the solution exhibits periodic behavior around  attractor in local domain,  same behavior obtained through dynamical analysis.
 \end{abstract} 
 

 \maketitle

\section{Introduction}
 The understanding of coupled multi component  dynamical  system (steady state or bifurcation) requires mathematical understanding of the system  manifold.
 In particular, generic bifurcation theory with symmetry, normal forms and unfolding theory all make vital contributions to explain and predict behavior in such systems.                                 
We consider pathogen dynamics model or immune network model where immune response in target invasion and proliferation in body system is considered to follow 
very non linear/complex path. The growth and interaction pattern follows non linear predator- prey type interaction. 
CD8+ T cells are one of the most crucial component of the adaptive immune system that play key role 
in response to pathogen. Upon antigen stimulation, naive CD8+ T cells get activated and differentiate
into effector cell. This mechanism may create small sub set of memory cell after antigen clearence.
Emerging evidences support that metabolic re programming
not only provides energy and bio molecules to support pathogen clearence but is also tightly linked to T - cell 
differentiation \cite{bevil}. In brief, naive CD8+ T cells are activated in response to the coordination
of three signals, including TCR, co stimulation and inflammatory cytokines via multi step strongly  connected complex pathway. The majority of CD8+ T cells undergo contraction phase and die by apoptosis \cite{bevil}.
Therefore, we try to undertake non linear pathogen dynamics mathematical model for present studies.
Since symmetry is fundamental  invariant structure associated with various mechanical /physical system, 
 it influences  functionality of the dynamical system.                                                                                                                 
Systems with hamiltonian dynamics, equation of motion exhibits symmetry in which total energy conserved (Noether Symmetry).  
On the other hand, various physical dynamical systems 
exhibit symmetry feature which conserves action of dynamics and gives rise to Euler - Lagrange equation as 
equation of motion. 
 Many dynamical systems represented by coupled autonomous equations exhibit presence of attractor 
(local or global) to sustain stability of the dynamics. 
 The work of Aswhin et al \cite{ ash} showed connection between transitive symmetry algebra and 
periodic behavior of dynamical system
which indicates reflection of pattern behavior  through 
existence of symmetry structure.  Symmetric attractor is signature property of equivariant dynamical system. 
 Continuous group such as compact lie  group is used in many mathematical models to understand connection 
 between symmetry and invariant quantity in the dynamics.
  Followed by such idea, we try to explore pattern behavior under action of continuous group symmetry. 
  Since Lie group action under one parameter point transformation leaves the manifold invariant (under linear vector field),
   this can be used to unfold evolution structure to obtain pattern structure at any time. 
 Moreover, maximal Lie algebra for a second order ODE can leave diffeomorphic manifold invariant, this can 
be implemented to integrate non linear ODE. 
  In most  autonomous equations, symmetry algebra is transitive in nature
 (group generators follow simple time translation and population growth). 
 Conn et al \cite{conn} described transitive Lie algebra over a ground field K (real or complex field) as  topological 
 lie algebra whose underlying vector space is linearly compact and which possesses a fundamental sub algebra with no ideal 
 (opposite to the case of primitive action algebra). 
 
From standpoint of geometric analysis of Lie algebra, generator  should take the  form 
\begin{gather} 
 \xi = \sum_j \xi_j \cfrac{ \partial}{ \partial t_j } \\
 \xi_j \in F 
\end{gather} 
which we view as formal vector field under Lie infinitesimal  transformation. 
Under  lie group of infinitesimal  transformation,  system follows connected manifold. 
Given a connected differential manifold M and the action of a compact Lie group g  on M, $ g_t  $ represents the isotropy sub group 
 of  $ \mathrm{g } $  at any time t for a dynamical system represented by autonomous ODE. 
Normalizer $ N( g_t) $ of the sub group must exist corresponding to $ \tilde{t} $.  If 
 g( t) and  $ g ( \tilde{t} ) $  are equivariantly diffeomorphic if  $ g_t $ and $ g_{\tilde{t} } $ are conjugate subgroup of g. 
  Equivalently, g( t) and $ g( \tilde{t} ) $ 
 are equivalently diffeomorphic.  If $\vec{v} $ is an equivariant vector field (group generator), then $ \vec{v} f =0 $, 
  where f represents differential equation  of the dynamical system.    
   Local pseudo group (local diffeomorphism) under transitive Lie algebra preserves structure of the manifold 
very much which is very important in stability of dynamics.  
    Theorem of Guillemin \& Sternberg \cite{gui} asserts that, given a transitive Lie algebra L and a fundamental sub algebra $ L^0 \subset L $, 
    one can realize L as a transitive sub algebra of formal vector fields in such a way that $ L^0$ is realized as isotropy sub algebra of L.;
     such realization of  $ (L, L^0) $ is very unique under formal change of coordinates. This means under group action 
(vector field $ \vec{v} $), 
     flow $ t \rightarrow exp( t \vec{v} ) $  is maintained.                                                                                                                                                                                            
 In most physical problems dictated by hamiltonian of the system (Kepler s law of planetary motion), infinitesimal transformation of the variables under Lie group of continuous transformation showed  momentum conservation (Noether Symmetry).
 In system dictated  by Lagrangian (action integral) of the system,  
 such group symmetry can manifest in obtaining some of first integral (under action of proper sub algebra) 
which can be related to Lagrangian of the system.  

  In this work, we construct evolution structure driven by presence of symmetry (Lagrangian or Hamiltonian). 
  Since biological evolution does not follow conservative system, one can assume the system is driven by Lagrangian action (followed by Euler - Lagrange equation).                                                                                                                                                                                                                                                                                                                                                                                                                                                                                                                                                                                                                                                                                                                                                                                                                                                                                                                                                                                                                                                                                                                                                                                                                                                                                                                                                                                                                                                                                                                                                                                                                                                                                                                                                                                                                                                                                                                                                                                                                                                                                                                                                                                                                                                                                                                                                                                                                                                                                                                                                                                                                                                                                                                                                                                                                                                                                                                                                                                                                                                                                                              
The stable structure of the dynamics is intrinsically connected to its inherent symmetry to the system. Once, we are able to obtain such symmetry, that is used to obtain analytical structure of the solution which  means irrespective of initial condition,
 the solution will follow same behavior globally/partially.  
The work by Ashwin \& Melbourne et al \cite{ash} 
 gave necessary condition for a subgroup of a finite group ( solvable sub algebra)  to have symmetry of a chaotic/non chaotic attractor.    
 Their numerical study  showed that if any solution of the dynamical system (described by ODE) is obtained from G invariant system of ODE ( PDE)
  where G is a compact Lie group of symmetries, then even if solution varies chaotically in time , there exists some invariant quantity under non trivial sub group of symmetries in G. 
 The observation of periodic behavior in solution can describe system having symmetry.  
 With such extensive work and results by several groups to understand existence of symmetric structure of a 
 dynamical system and corresponding regular pattern of solution, we try to investigate immune dynamics model to study
  relation between existence of symmetry and solution structure.  Our goal is to study possible existence of symmetry algebra 
  of continuous group under invariance and use that to develop analytical structure of solution, if it exists. 
 In this context, we discuss 
methods to construct vector fields that act continuously on variables on diffeomorphic manifold 
(under action of compact Lie group \&  transitive algebra).
  Since  Lie group  with maximal symmetry can have  canonical coordinates under converts ODE  into 
 quadrature form, it will be easy to obtain solution in linear sub space.  
 Once solution structure is obtained, one can add non linear term into linear part  to study effect of non linear term. 
We adopt group theoretic approach to construct symmetry structure and differential invariant
 of the manifold under action of G. 
 Our goal is to find existence of Lie symmetry obeyed by ODEs conditionally or globally
if it exists and then use it systematically to obtain analytical structure of the solution. 
Since Lie point symmetry vector field is linear, we can assume under action of the symmetry group, 
system preserves dynamics  of least path of action or action of Lagrangian is preserved. 
Under Lie point group action g, if the dynamical system exhibits convergence properties , then 
system can contract towards local fixed point. 
The characteristics of many physical dynamics is that system  possess intrinsic symmetry characteristics of certain kind of conservation law under symmetry algebra. 
 First integral  method is very significant as first integral is associated with Lagrangian of the dynamics. 
First integral may confine the solution to a bounded region of phase space.
 It is very important to have knowledge about analytical structure of solution to understand the dynamical system. 
 Certain mathematical community have devoted their research on algebraic structure of various point symmetries in various
dynamic system.
 
   In order to obtain analytic solution of the ODE through symmetry, 
method involves reduction of order through construction of canonical ( normal ) sub space using solvable sub algebra.  
Derived algebra of a Lie algebra G is analogous to commutator subgroup of a group. 
It consists of all linear combination of commutators and clearly an ideal.  
 For a r dimensional Lie algebra $ G^r $, relation is given by
\begin{gather} 
 [ g_a , g_b] =  {\sum_{ \gamma=1} }^r  {C^{\gamma} }_{ a b} g_c 
\end{gather} 
in terms of structure constant $  {C^{\gamma} }_{ a b} $.
 $ g^r $ is r dimensional solvable algebra if there exists a chain of sub algebras 
 \begin{gather} 
 g^{(1)} \subset L^{(2)} \subset \cdots g^r = g^r 
 \end{gather} 
 such that $ G^{(k)} $ is a k dimensional Lie algebra  and $ g^{(k-1)} $ is called an ideal of $ g^{(k)} $.  To obtain solution in differential invariant sub space, it must satisfy 
 \begin{gather} 
 [g_a, g_b ] = \lambda g_a  
 \end{gather} 
 for any positive  $ \lambda$. 
  First method involves docntsruction of normal form of generators in linear sub space of canonical variables and invertible mapping. 
 Under solvable sub algebra, normal form of generators convert ODE into quadrature form. 
  The second method involves construction of differential invariant function in the space of invariant function which can then be utilized to construct solution.  This method requires kth extended generator formation  of an ODE.
  
Any equivariant dynamical system possesing  Lagrangian/hamiltonian  structure or any kind, should possess recurrent robust attractor \cite{ash}.  
Since symmetry plays fundamental role in many physical/mathematical problem, our main focus will be  to develop condition for  symmetry that drives immune network.   
Many systems in nature posses 
intrinsic dynamical symmetry. 
We can consider biological evolution system to follow  Lagrangian dynamics  where action as functional drives evolution mechanism.  In many physical systems, 
  Lagrangian function remain invariant associated with the  symmetry. 
  It can be assumed  to have rich interplay between symmetry property and dynamical behavior. 
The experimental work of Ma et al \cite{ma} showed periodic behavior of infection phase of a patient in rubella infection.  Since most of the clinical data takes 
average data from blood sample the severity of the disease, it can not reflect the detal dynamics in infection and chronic phase of the disease. The clinical data by Liao et al \cite{liao} and references therein  in 18 such pathogen borne infection cases suggest periodic nature of the infection and related symptoms. This means fever or other external symptom 
follows up-down path over time till it disappars finally or requires external intrvention to annihilate target proliferation.
All these clinical result support complexity in interaction path.  Keeping complex nature of immune -target network, we consider non linear autonomous equation for pathogen dynamics in next section.

 Following is our plan of work: \\
  Section I is the introduction. Section II describes  basic immune dynamics model  with various features and 
 section III describes detail dynamical analysis of the model.
 In section IV \& V, we construct method of fundamental symmetry generators under
 infinitesimal transformation (under transitive algebra). 
Results of numerical simulation is also shown using group theoretic structure of the solution in last section. In this work, 
we try to understand the pattern of growth behavior and corresponding interaction phase space.

 \section{ Immune Dynamics Model} 
 
  In  case of any taget invasion or infection (bacteria/virus/immnulogic tumor cell ) in body, two types of  immune cells, namely 
 effector and memory cells play key role  as  immune response in combating such infection in short or long term. 
The proliferation and interaction of target cell in body is multi component/step  non linear phenomena following idea of predator- prey dynamics.
 The dynamics can be represented as 
 \begin{gather} 
 \dot{x}_i = \sum_i c_{ i j} x_j - x_i \sum c_{ k j} x_j 
 \end{gather} 
 where  first term designates self proliferation and second term mutual interaction. 
In case of major two component pathogen dynamics, system is described \cite{mayer}  as
   \begin{gather} 
\dot{x}(t) = r x(t) - k x(t)  y(t) = F( x, y) \\
\dot{y}(t)  =  f(x) + g(y) - d y(t) = G( x, y) \\
f(x) =  \cfrac{ p x^u}{ m^v + x^v}  \\ 
g(y) =  \cfrac{s  }{ c + y }y(t) 
\end{gather}
where x(t)  \& y(t) represent target cell and  immune cell density in local volume at any time t.  
 The term f(x)  represents velocity of  immune stimulation by target invasion x(t=0 ) 
which leads to competence against them in the network  with  u and v  as degree of stimulation, respectively. 
 The form of f(x) and g(y) represent Richardson type 
logistic functional growth in model of competence. 
The term g(y)  represents  autocatalytic enforcement of the network which is very necessary to acquire adaptive immune memory cell infection, as manifestation of chiral autocatalytic origin. 

The parameter m represents  threshold presence of target cell that can be recognized as signal by immune network.  Parameter 
  d represents constant  death rate of immune cell.  Immune competence y(t)  can be  defined as elimination capacity of the immune system  with respect to 
  target \cite{mayer} and can be measured by the concentration 
 of certain cytotoxic T - cell, natural killer cell or by concentration of certain antibodies.                                                    
Cross talks between antigen presenting and T cell impacts cell homeostasis amid bacterial infection and tumorigenesis \cite{ gau} dynamics.   
The condition u= v is characterized by no target burden factor or equivalently target and immune cell growth are in competence \cite{mayer}. 
We chose this non linear model of infection disease in our studiies of role of symmetry algebra and 
how to obtain analytical solution structure. 	
This type of growth pattern of immune  cell is noticed  in tumor dynamics and other pathogen infection  \cite{ mitra}.

\section{ Dynamical Analysis of the Model} 
Through dynamic analysis ,  we try to study  robustness of this kind  model in terms of stable invariant phase space.
												
Here,  we implement  no effective immune cell present at t=0 as a trigger effect of  f(x)  and study  immune netwrk growth pattern. 
 Following stability analysis, we obtain one equilibrum point  of  zero  target and positive nonzero immune cell with  following relationship of essential parameters
  which work as basin of attraction of the two component dynamics , namely 
 $ \left( 0, \cfrac{ d( c d -s) } {s^2} \right)$ with
 \begin{gather} 
m = - \cfrac{ k p ( c d - s) } {d r}  
 \end{gather} 
 Here, m in model represents threshold value of target population triggers immune network. 
 Corresponding eigenvalues of Jacobian are evaluated as 
\begin{gather} 
E_J  = 0, \cfrac{ d( c d -s)}{s^2}
\end{gather} 

 \vspace{1.0in}
 Dynamic simulation data ( FIG 1 \& 2)   exhibit correlated dynamics between two components in the network through production of so called proper antigen mechanism path.  \\

 Linear dynamics endowed by u=v=0 shows the pattern behavior in both component in closed phase trajectory (FIG 1) 
around point of attraction. 
 \begin{figure}

\centering
\subfigure{
\epsfig{file=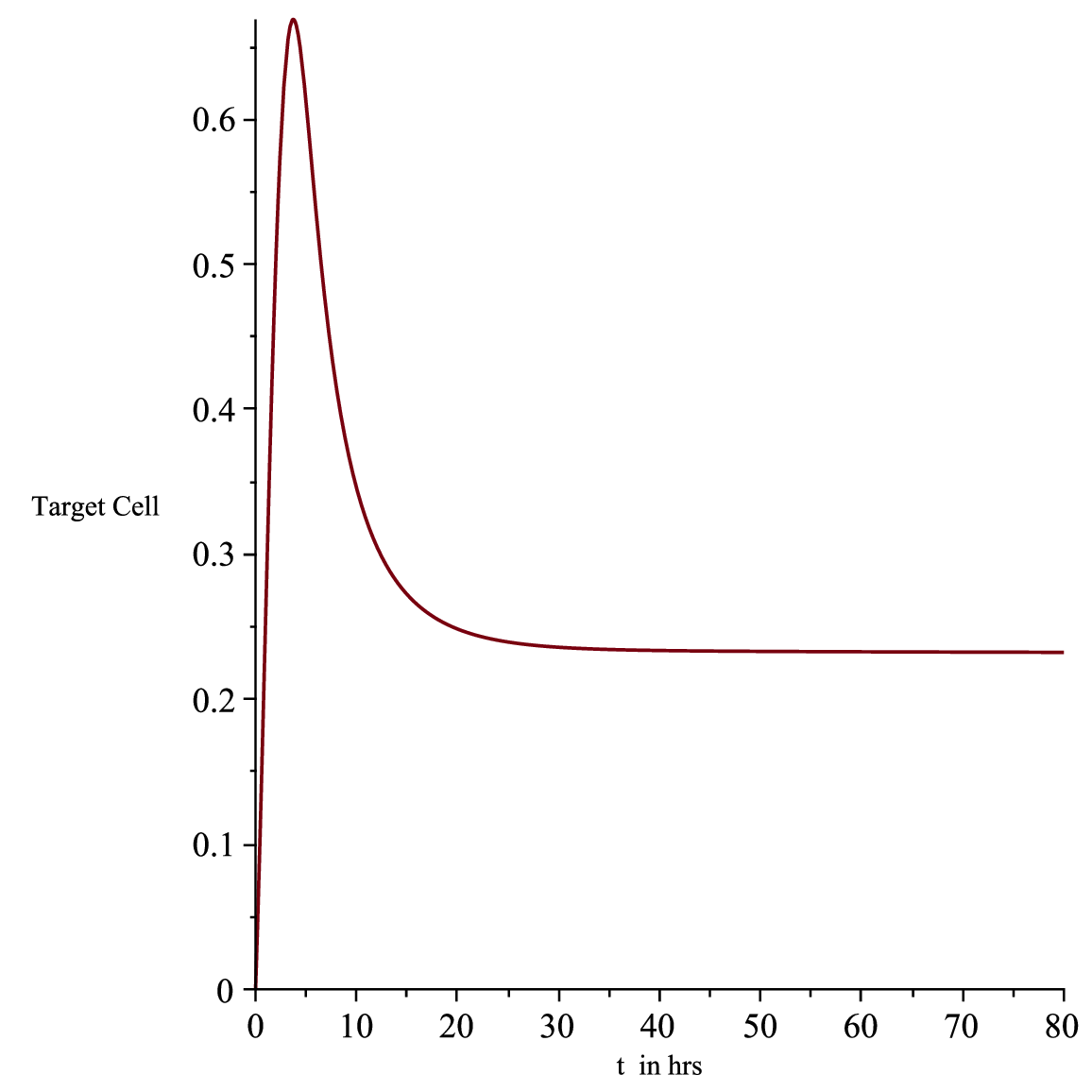, scale=0.25}}
\hspace{8pt}
\subfigure{
\epsfig{file=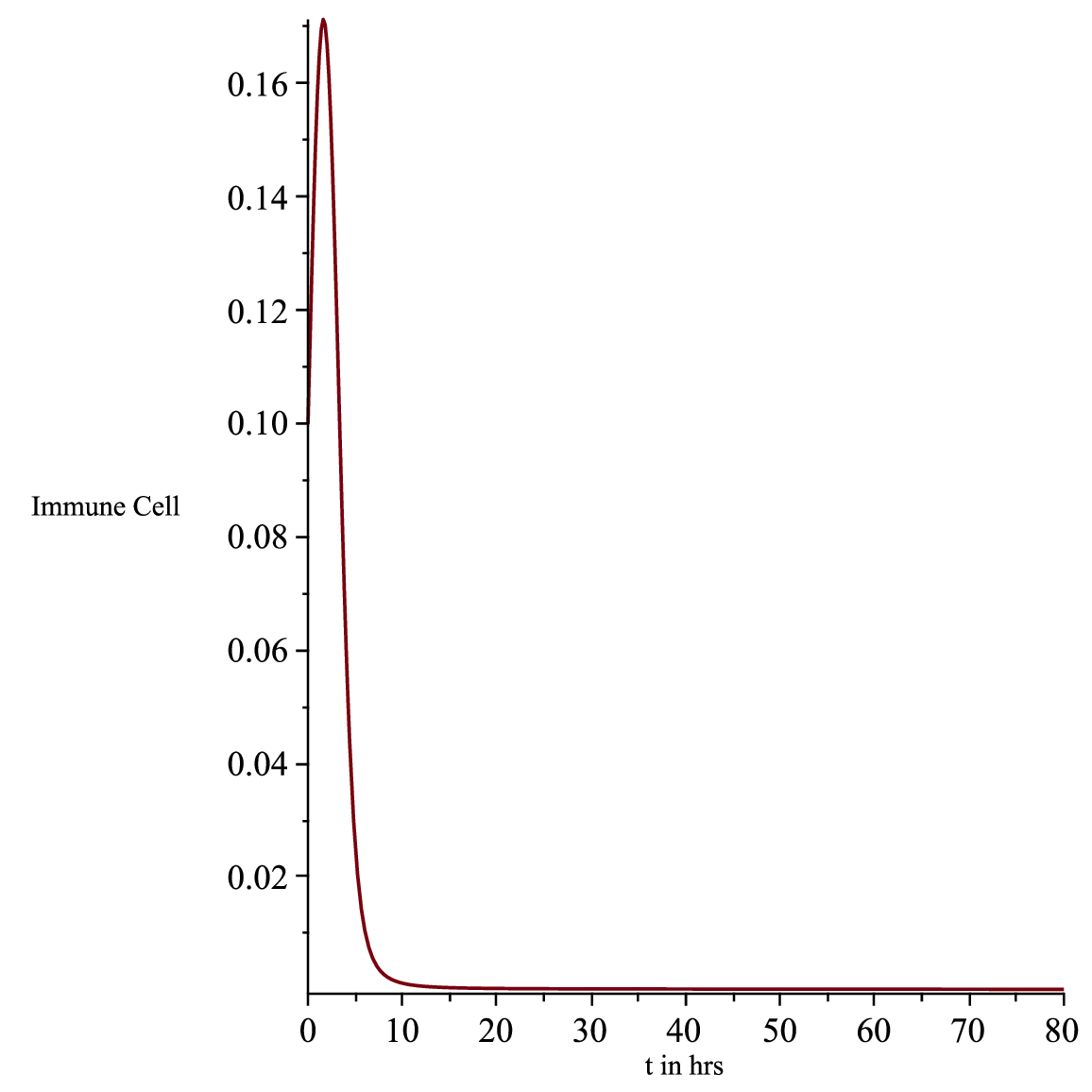, scale=0.25}}
\hspace{8pt} 
\subfigure{
\epsfig{file=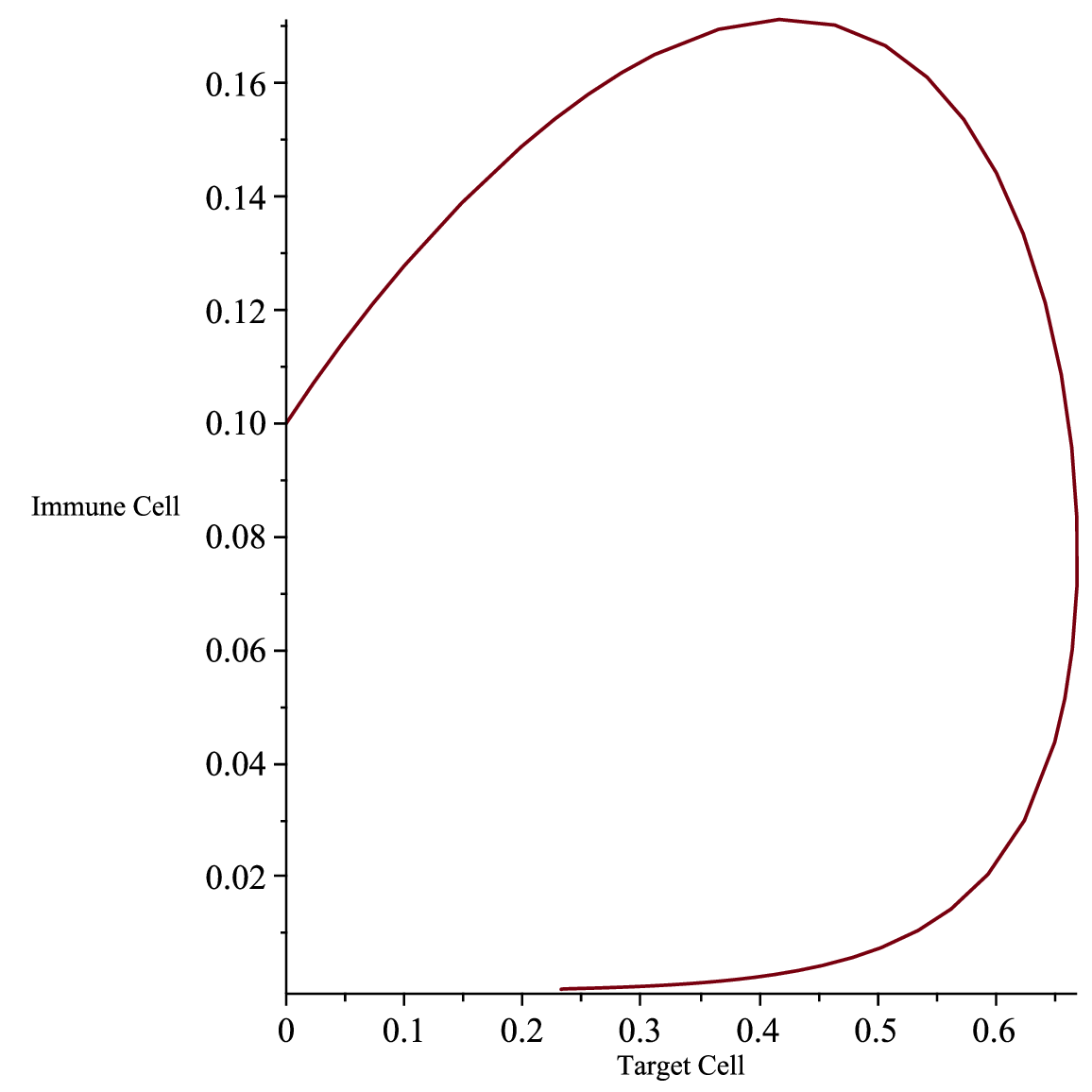, scale=0.25}}
\caption{Linear Growth Pattern;   d=1.0, r=0.701, p=0.642, s=1.23, c=1.0,  u=v=0, k= 0.9255, m= -0.14 } 
 \end{figure}    

 \vskip 4pt 
Adding non linearity into simulation setting u=v=1,  we increase s (velocity rate of immune cell interaction) to 1.53, the system exhibits oscillation phase even target interaction rate k is low. 
This is shown in Fig. 1 \& 2 where very stable phase trajectory is exhibited. 
The coexistence of stable pattern of both component is illustrated in Fig. 2. The dynamics follows limit cycle characteristics
with negative value of m showing equilibrium point.

\begin{figure}
\centering
\subfigure{
\epsfig{file=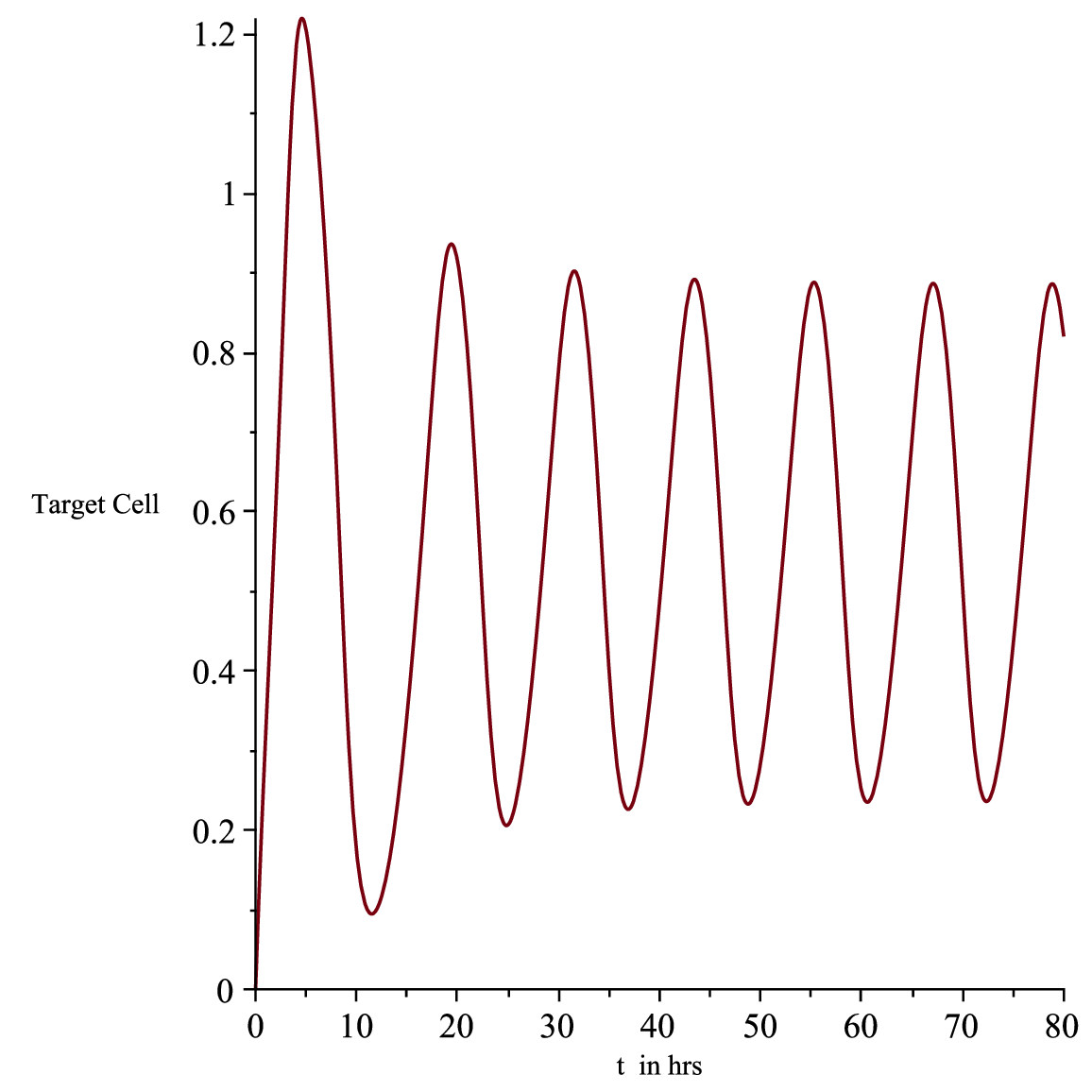, scale=0.25}}
\hspace{8pt}
\subfigure{
\epsfig{file=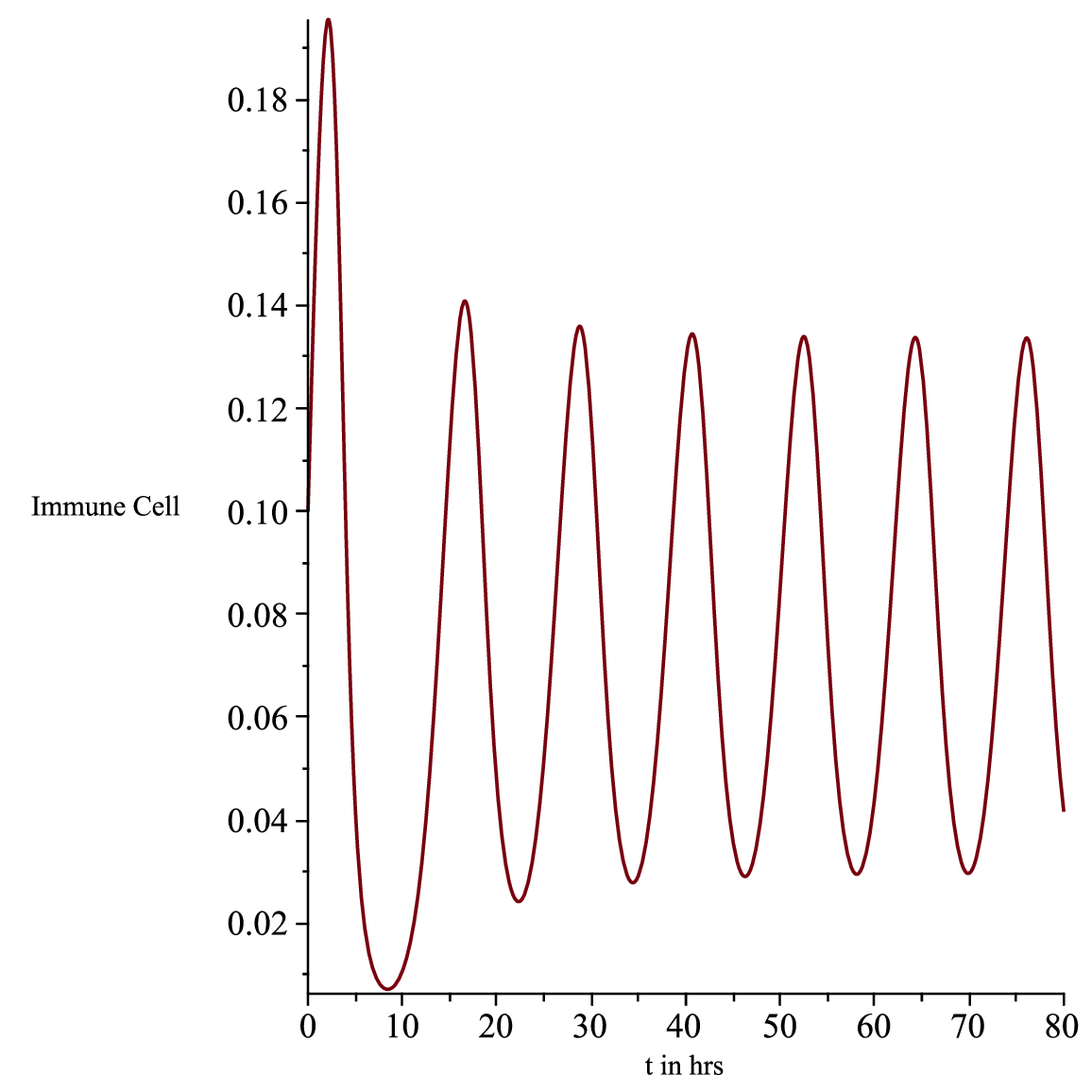, scale=0.25} }
\hspace{8pt} 
\subfigure{
\epsfig{file=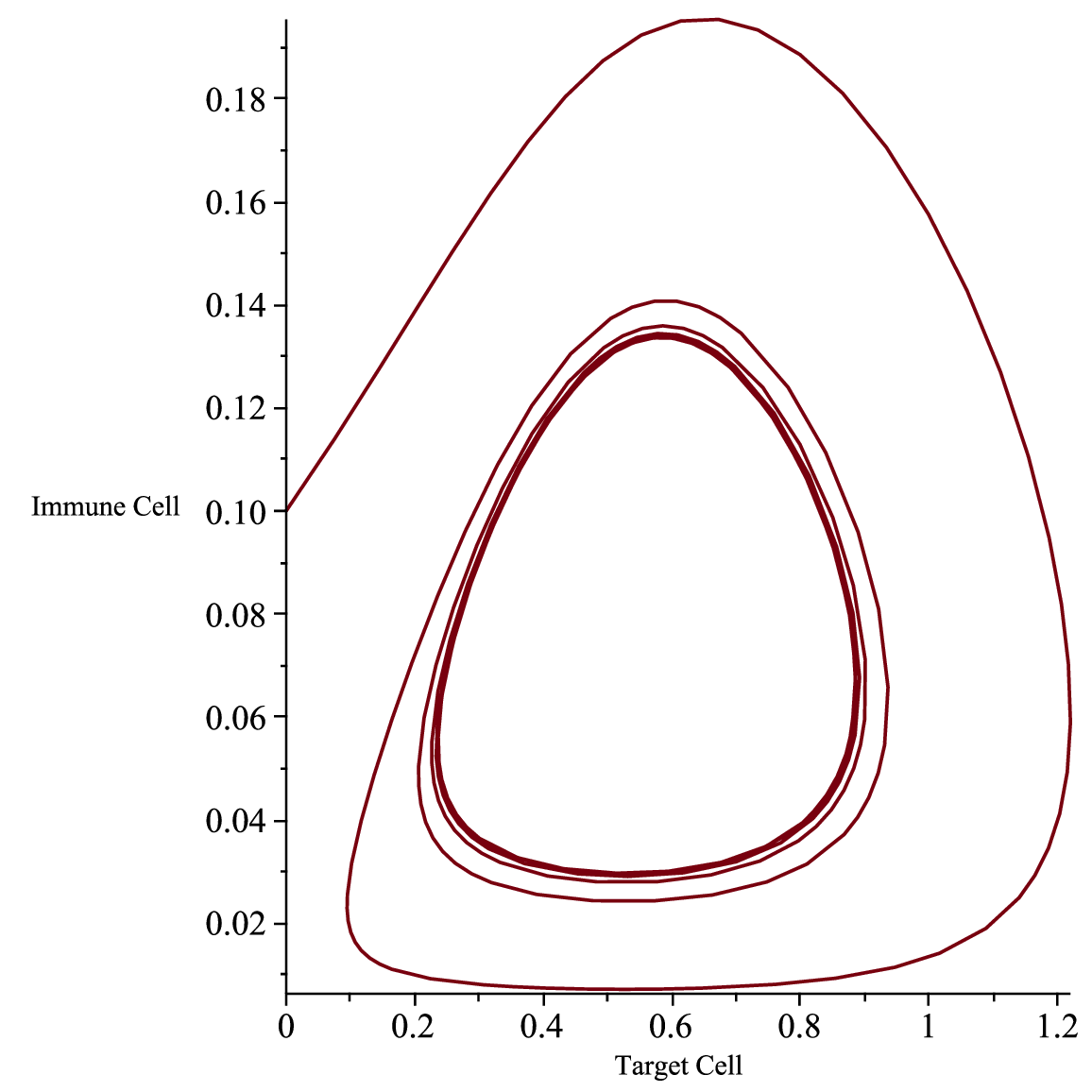, scale=0.2}}
\caption{Periodic behavior of growth dynamics  ;   d=1.0, r=0.701, p=0.642, s=1.56, c=1.0, u=v=1, k= 0.12, m= - 0.12  } 
 \end{figure}    
 Our analysis reveals significant role of m in determining stability of the dynamics ( $ - 0.1 < m < 0.3 $ ).
When value of m shifts from this range in either direction, co existence of stable pattern disappears. 
The oscillation becomes more prominent and stable for longer time when we set v=3 for negative value of m shown in  FIG 3. 

 \begin{figure}
\centering
\subfigure{
\epsfig{file=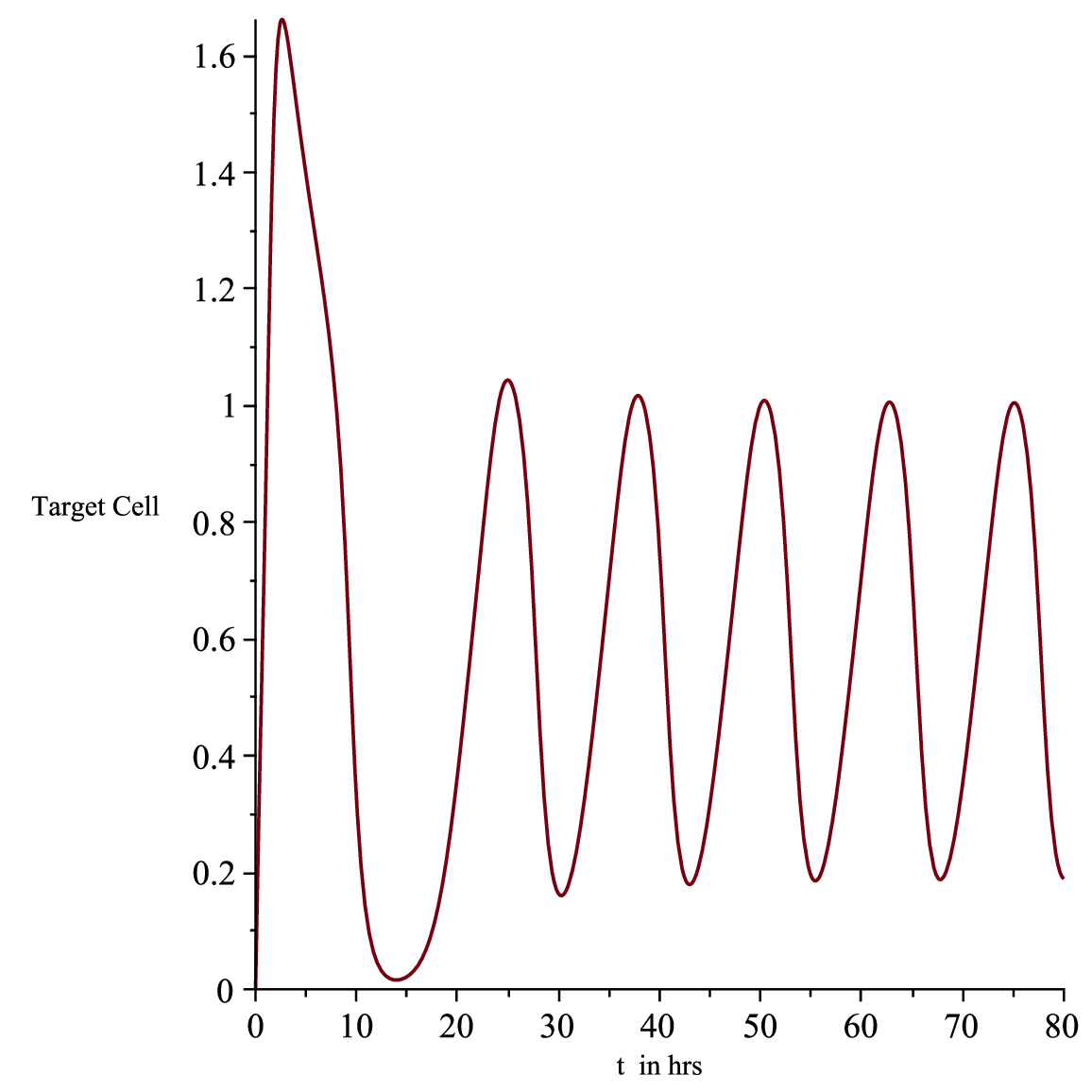,scale=0.25}}
\hspace{8pt}
\subfigure{
\epsfig{file=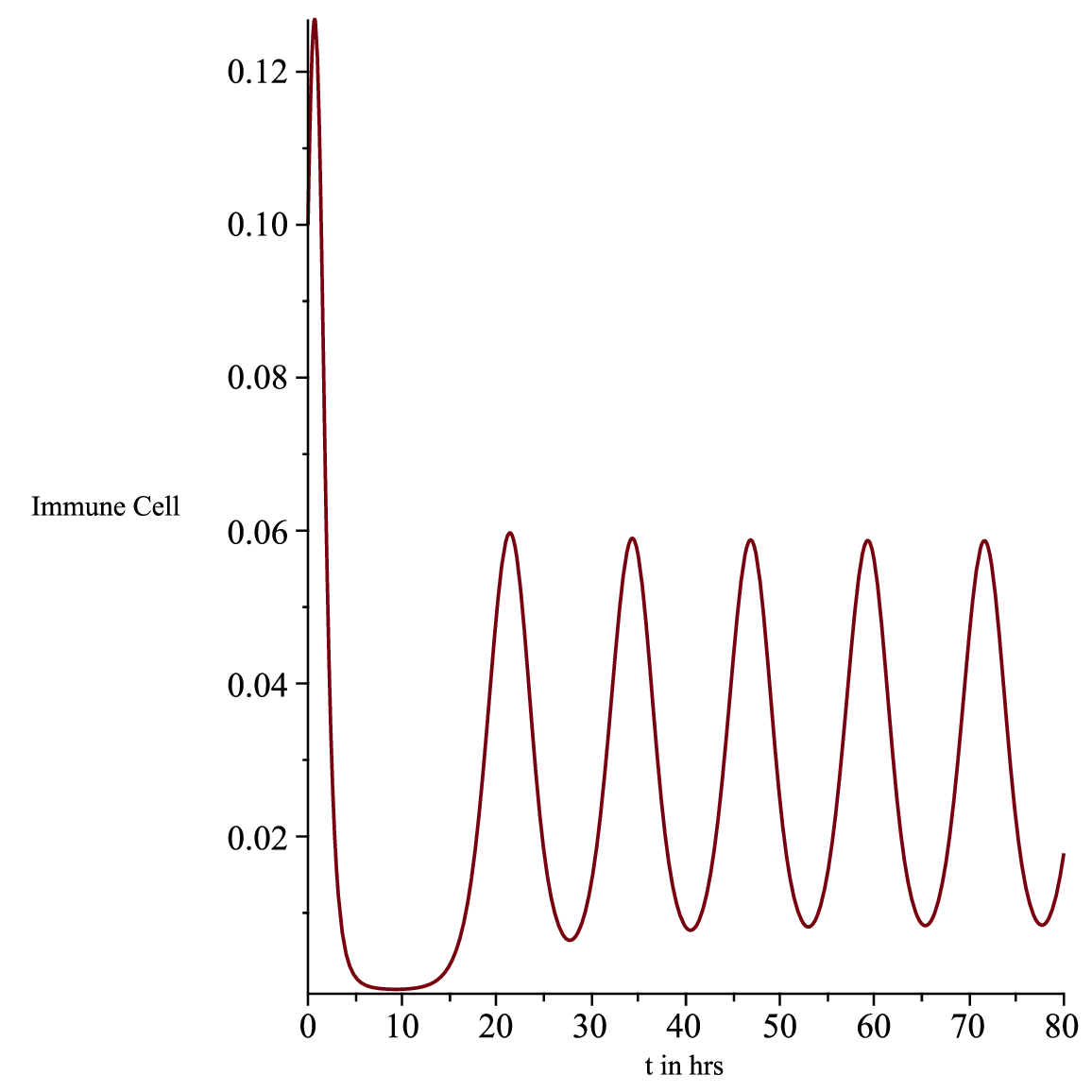, scale=0.25}}
\hspace{8pt} 
\subfigure{
\epsfig{file= 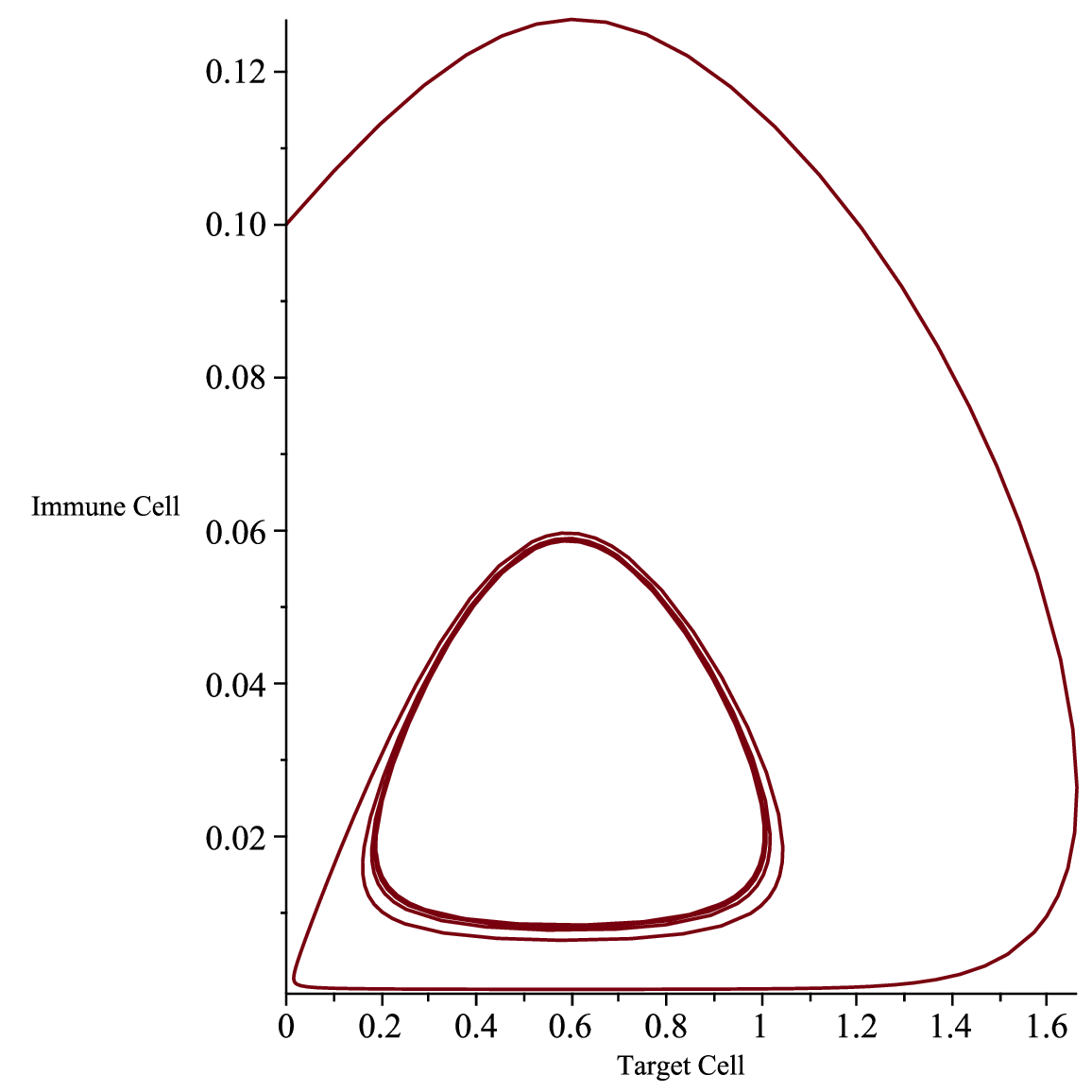, scale=0.25}}
\caption{Periodic orbit presence in correlated dynamics  ;   d=1.0, r=0.701, p=0.642, s=1.56, c=1.0, u=1, v=3, k= 0.12, m= -0.9953  } 
 \end{figure}

 This regime can be recognized as interacting phase of  various T/B cell via cytokinase production mechanism in order to achieve immunity for longer period marked by m =- 0.9953.
 Numerical data exhibits characteristic pattern change after 9-12  hours of the cell infection. 
  Fig. 2 \& 3     exhibit characteristic pattern  with  stable closed  trajectory around local attractor. 
    This phase can also be termed as long term antibody production phase and  is significant in rubella, german measles, 
 influenza, smallpox etc or any other lethal disease. Oscillation occurs  in presence of  threshold  target cell ($ x_c  + m $) recognized by immune network.
 This is the case  immune system recognizes small presence of target cell through special 
 kind of T cell presence in the system. Existence of multi layer component in immune network in case many lethal 
diseases such as rubella, HIV are very common. 
It is evident that  ratio of target proliferation $ \cfrac{  p x^u}{ x^v} $ is very dominant  to sustain pattern coexistence. 
  Upon increasing a u and  v  in the network, we observe existence of chaotic phase trajectory  shown in FIG 4. 
  
  \begin{figure} 
\centering
\subfigure{
\epsfig{file=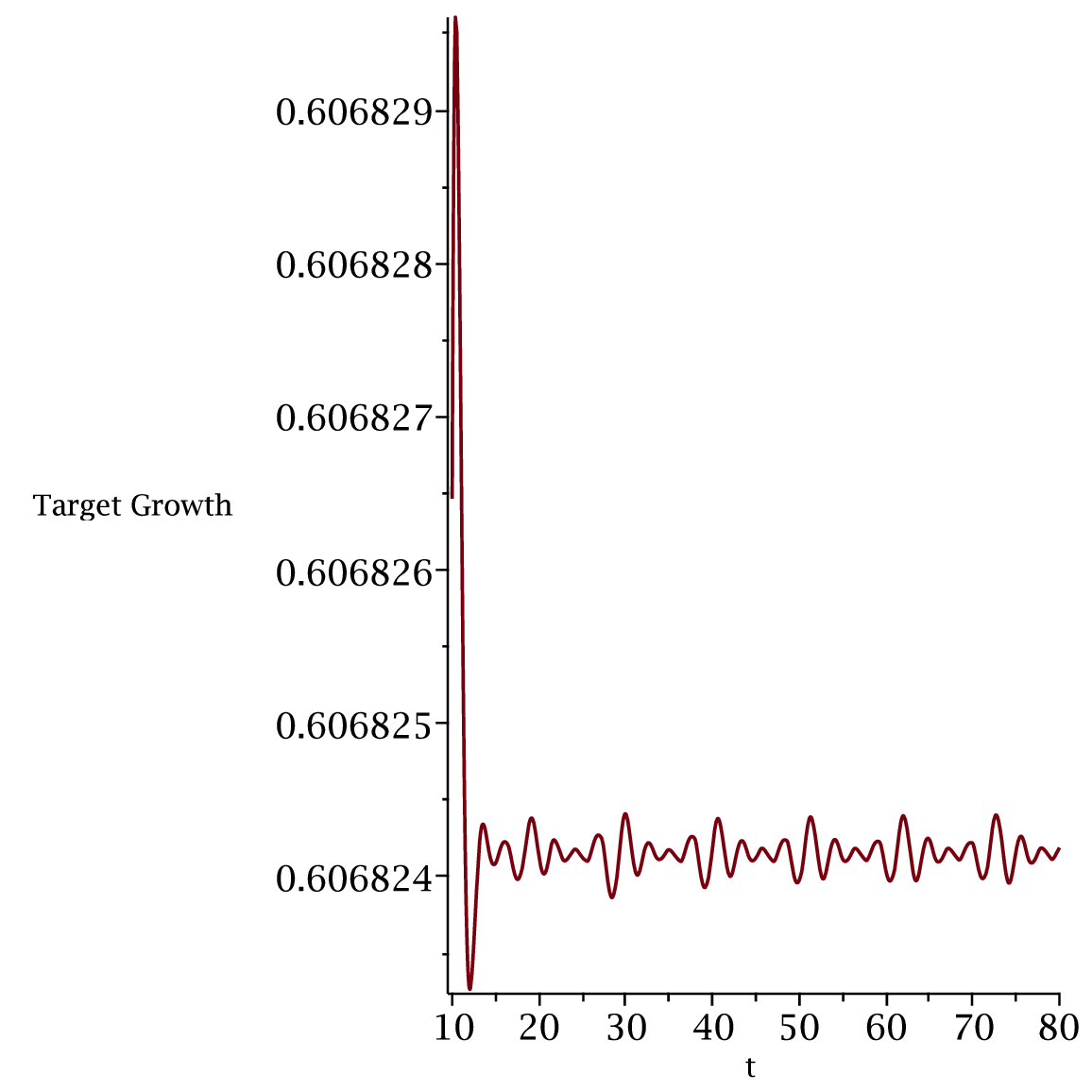, scale=0.25}}
\hspace{8pt}
\subfigure{
\epsfig{file=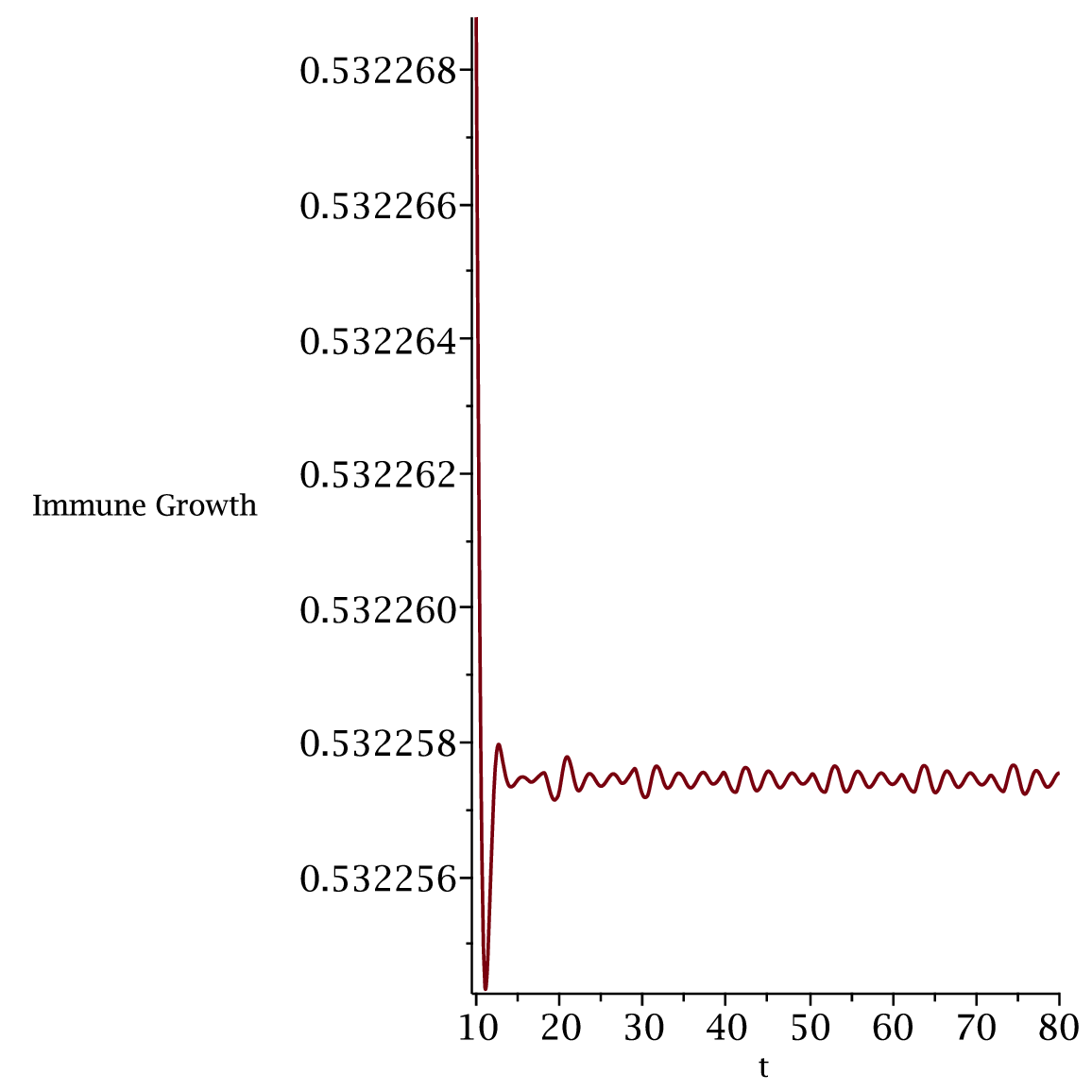,scale=0.25}}
\hspace{8pt} 
\subfigure{
\epsfig{file=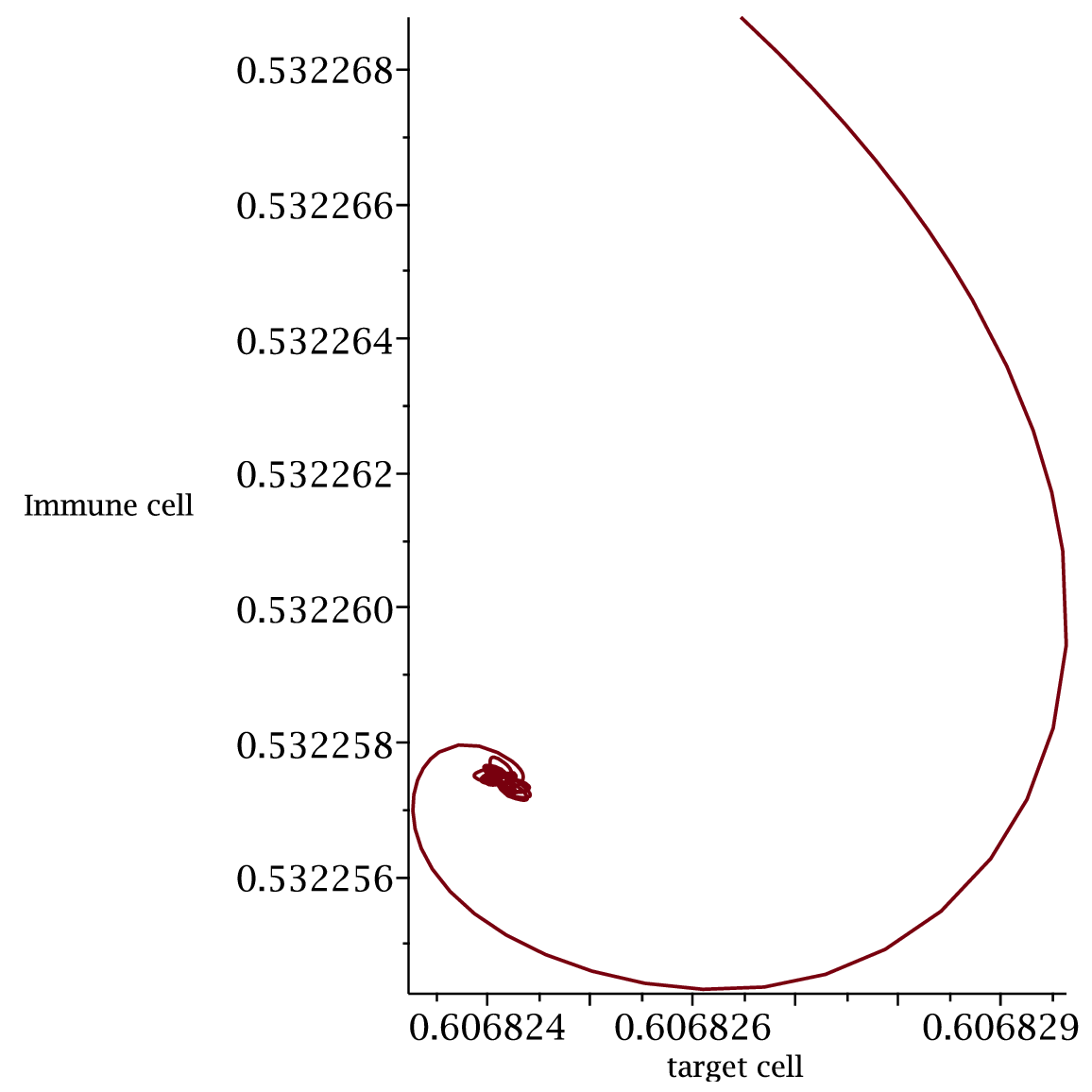, scale=0.25}}
\caption{Chaotic phase trajectory ;   m = 0.9995 ,  d=1.0, r=0.701, p=0.642, s=1.72, c=1.0, u=2, v=4, k= 0.265 } 
 \end{figure}    

  This kind of oscillatory/chaotic  behavior can be termed as indeterministic dynamics  where solution can not be obtained  following deterministic methods.
Stochastic  variability of target concentration or strain type  within a period of time gives rise to such dynamics.

Moreover,  we observe  target cell growth becomes zero with very sharp increase of immune cell  for $ m > 0.25 $ . 
This regime  can be termed as  antibody pattern recognition by immune network shown  in FIG 5.  
 
\begin{figure} 
\centering
\subfigure{
\epsfig{file=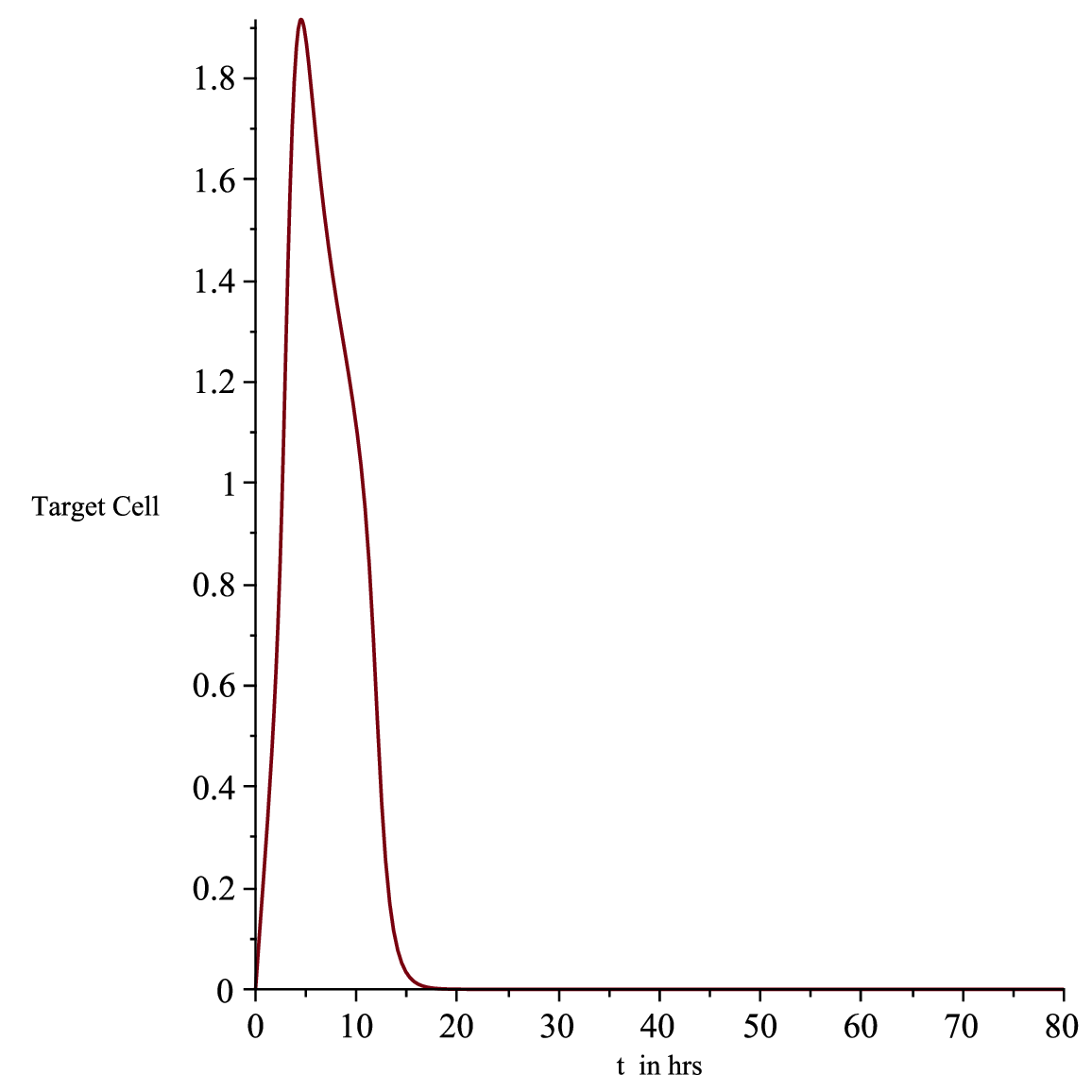, scale=0.25}}
\hspace{8pt}
\subfigure{
\epsfig{file=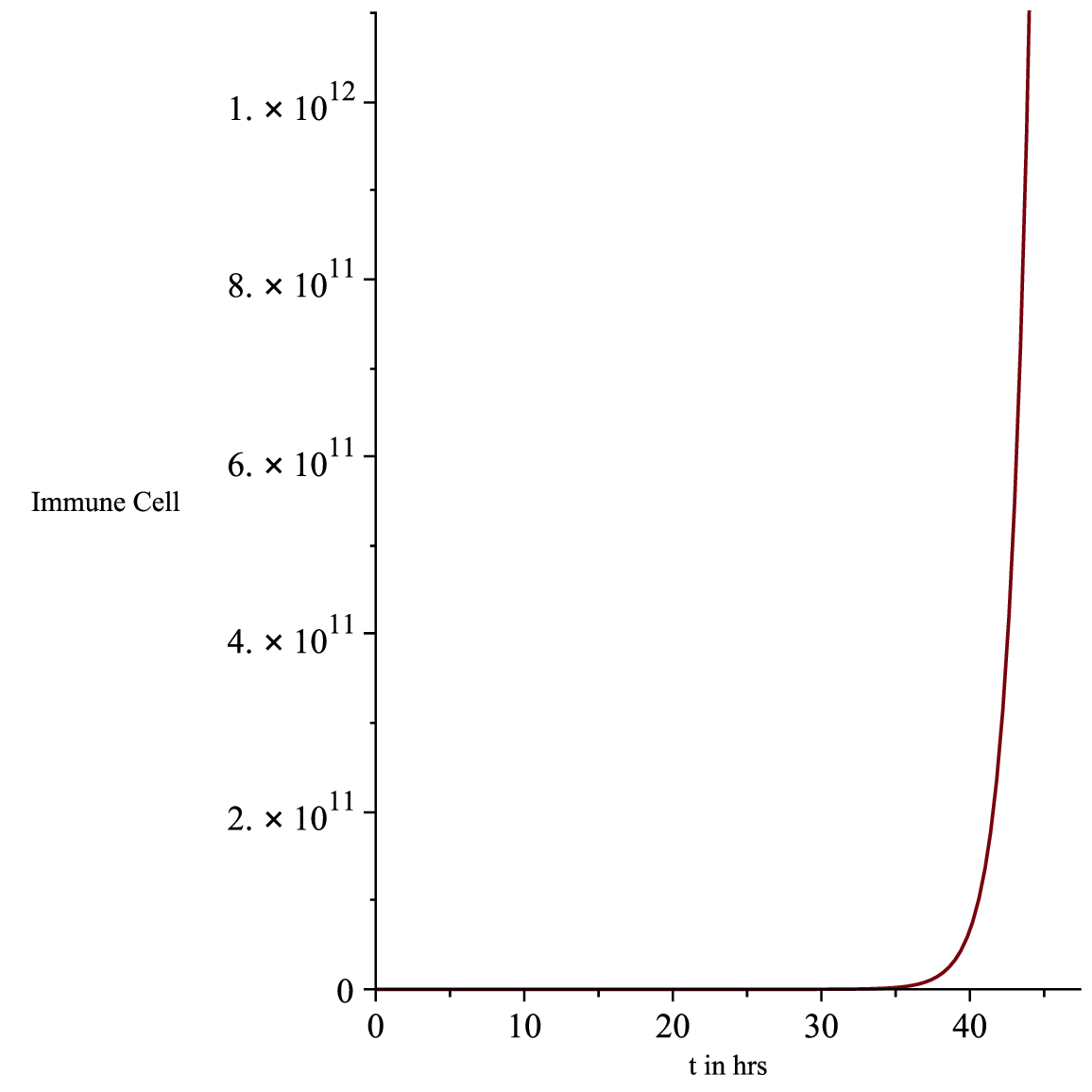,scale=0.25}}
\hspace{8pt} 
\subfigure{
\epsfig{file=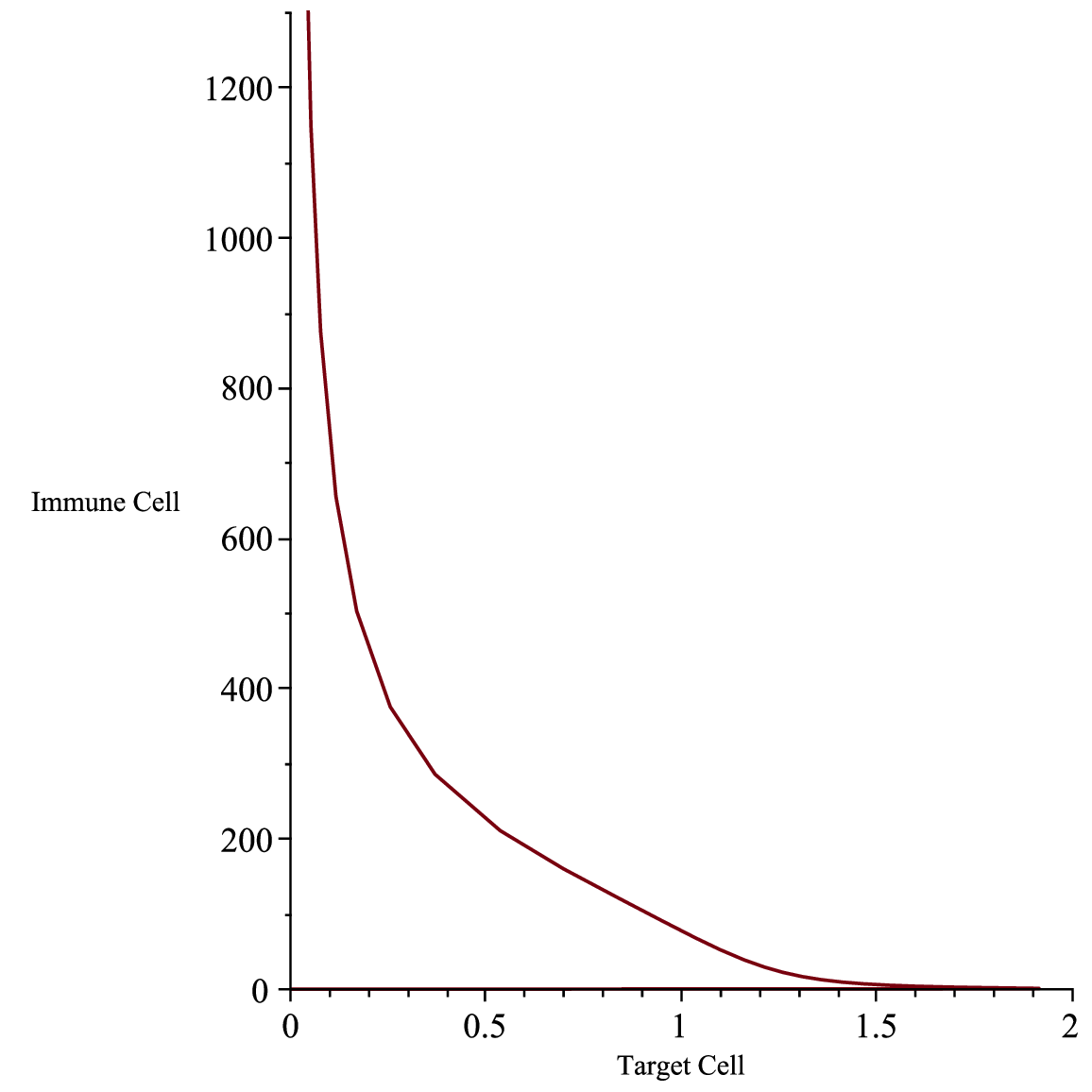, scale=0.25}}
\caption{Behavior far away from critical attractor   ;  $ m \ge 0.3$,  d=1.0, r=0.701, p=0.642, s=1.72, c=1.0, u=v=1, k= 0.265 } 
 \end{figure}

The  open region  in FIG 5   is marked as  therapeutic intervention case, often  recognized by  
  poor/failed immune system and  physics can be explained via  uncorrelated/random response of T cell in lymphocyte.

\section{Transitive lie algebra method to obtain analtytic solution structure} 

In order to study symmetry algebra and its role on dynamics, 
we plan to  obtain invariant lie symmetry generator based on  Lie symmetry algebra in manifold. 
The dynamical system
\begin{gather} 
\ddot{x} = f(t,x)
\end{gather}
will be  equivariant under  Lie group  $ \mathrm{g} $ on  m (m=2)  dimensional manifold M  with symmetry group of operator $ g^i  $, 
then following must be true under periodic temporal symmetry, i.e 
  \begin{gather}
  f(\vec{g}^i ,  x ,t) = {g}^i  F( x, t + t^{\prime} )  
  \end{gather} 
  for any  $ t^{\prime}  > 0 $ and the system is called $ \mathrm{g} $  symmetric. 
   
  In this case, $ \mathrm{g} $   acts locally on  m - dimensional manifold M. We are interested in the action of 
  $ \mathrm{g} $  on p - dimensional sub manifolds $ N \subset M $, which 
  we identify as  graph of functions in local coordinates.  The symmetry and rigidity properties of 
 sub manifolds are all governed by their differential invariants.  Here  group $ \mathrm{g} $  acts continuously on the differential equation. 
The model considers   target invasion with growth rate r  and  appropriate immune network response given by function g(y) and f(x).
If above   dynamical pair of  equations ( 7- 8)  undergo evolution following path of symmetry , it is then necessary to obtain symmetry structure to understand the dynamics. 
 Lie point symmetry and path of transformation is well known to preserve certain dynamical invariant which can be associated with equation of motion, if it exists. 
 Since Lie compact group of continuous point transformation ( continuous group) is special linear group, it  can be related to dynamics of holonomic constraints. 
We assume immune cell responds in a minimal non linear way with  n=2 and target stimulation parameter   u=v=1. With this assumption, we  convert coupled  
autonomous equations into second order non linear  ODE in x(t) as
\begin{gather} 
- \cfrac{ \ddot{x}(t) }{ k x(t)}  + \cfrac{ {\dot{x}(t)}^2 } { k x(t)^2 } = \cfrac{ p x(t)} { m+ 1} - \cfrac{ p x_0} { m+1}  - \cfrac { x(t)} { {(m+1)}^2} + \cfrac{ p x_0} { { (m+1)}^2}  \\
+ \cfrac{ sr}{ ck} - \cfrac{ s}{ ck} \cfrac{ \dot{x}(t)} { x(t)}  - \cfrac{ s r^2} { c^2 k^2} + \cfrac{ 2 s r}{ c^2 k^2} \cfrac { \dot{x} (t)} { x(t)} - \cfrac{ s}{ k^2 c^2} \cfrac{ {\dot{x}(t)}^2 } { {x(t)}^2} - \cfrac{ dr}{k} 
+ \cfrac{d}{ k} \cfrac{ \dot{x} (t)} { x(t) } 
 \end{gather}
which can be expressed as non linear ODE 
\begin{gather} 
\ddot{x}  = f^k( t, x(t)^k 
\dot{x}^k (t) ) \\
 = \alpha_2 ( t, x) {\dot{x} }^2 +   \alpha_1( t, x) \dot{x}   + \alpha_0(,x) x(t) 
 + \cdots 
\end{gather} 
with parameters defined as 
\begin{gather} 
\alpha_0 =  \cfrac{ s r^2} { k c^2} - \cfrac{ s r} {c} - \cfrac{ p k x_0}{ m0^2} x^2 + \cfrac{ dr}{k}  \\
\alpha_1 =  ( \cfrac{s}{c} - \cfrac{ 2sr} { k c^2} - d) \\
\alpha_2 = ( 1 + \cfrac{s}{ k c^2} ) \cfrac{1}{ x} 
\end{gather} 
with $ x_0$ as initial value of x(t). 
ODE (18) is second order differential equation,  non linear in $ \dot{x} $ and may be linear/ non linear in x. 
The idea of linear form of $  f_t ( t, x, \dot{x} ) $  and related to infinitesimal transformation  that would entail  dynamical extravagance. 
Once, analytical structure of the solution is obtained, non linear term can be added  into solution. 
 
Through $ sl(3, \mathbb{R}) $ algebra, any two 
non commuting, non proportional symmetry generators should follow 
\begin{gather} 
g_a = \xi_a \cfrac{ \partial} { \partial t} + \eta_a \cfrac{ \partial} { \partial x} \\
g_b  = \xi_b \cfrac{ \partial} { \partial t} + \eta_b \cfrac{ \partial} { \partial x}  
\end{gather}
with
\begin{gather}
[ g_a, g_b ] = \lambda g_a 
\end{gather}
For any physical/biological  dynamics represented by second order ODE, differential invariant function of the system is connected to the Lagrangian which describes : 
 A non singular Lagrangian admits  a symmetry group having dimension  $\le 3 $ . A non singular nth order, $ n \ge 2 $ 
 admits a symmetry of  dimension  $ \le n + 3 $ \\
  If $  \mathcal{L} ( t, x^n ) $ is corresponding g  invariant Lagrangian with non vanishing Euler - Lagrange expression  $ E(\mathcal{ L} )$, then every g - invariant evolution equation should satisfy
 \begin{gather} 
 \dot{x}  = \cfrac{\mathcal{ L}} { E(\mathcal{L})} I 
 \end{gather} 
 where I is an arbitrary differential invariant of the group. 
   If $ \mathrm{g} $  represents  generators of solvable algebra, corresponding differential invariant can be constructed. 

  Under one parameter infinitesimal transformation of coordinates (continuous map), we can write
\begin{gather} 
\tilde{t}   = t + \epsilon  \xi ( t, x)  \\
\tilde{x}  = x + \epsilon \eta ( t, x)
\end{gather} 
in the neighborhood of identity 
with vector field defined 
\begin{gather} 
g  =  \xi ( t, x)  \partial_t  + \eta ( t, x) \partial_x 
\end{gather}
 for any $ 0 < \epsilon \ll 1 $.  The computation of the flow generated by the  vector field is often referred to as
 exponentiation of the vector field 
 under transitive algebra so that  each vector  in  group  $ \mathrm{g}$ can be integrated  through origin in $ \mathbb{R} $ .  
 It contains an open neighborhood of the origin and flow. 
 \begin{gather} 
  t \rightarrow exp( t g ) 
  \end{gather}
 is  smooth for any vector field.  In case of real field $ \mathbb{R}^m $,  constant vector field defined by $ g_a = \sum a^i \cfrac{ \partial} { \partial x^i} $ with 
 $ a = ( a^1, a^2, \cdots a^m) $ exponentiates to the group of translation
 \begin{gather} 
 \exp( \epsilon g_a) = x + \epsilon a 
 \end{gather} 
with $ x \in {\mathbb{R}}^m $. 
  Under  continuous symmetry group represented by  (26 - 30), 
  mono parametric group  generate  essential generators for $ r \le 8$. 
Equation (18 ) under infinitesimal transformation  yields 
\begin{gather} 
\dot{\tilde{x} } = \dot{x} + \epsilon ( \eta_t + ( \eta_x - \xi_t) \dot{x} - \eta_x {\dot{x}}^2 \\
\ddot{\tilde{x}}  = \ddot{x} + \epsilon (( \eta_x - 2 \xi_t ) - 3 \xi_x \dot{x} ) \ddot{x} \\
+ \epsilon ( \eta_{t t} + ( 2 \eta_{ x t} - \xi_{ t t} ) \dot{x}  \\
+ ( \eta_{ x x } - 2 \xi_{ x t} ) {\dot{x}}^2 - \xi_{ x x }  {\dot{x} }^3 
\end{gather} 
  for all  $ ( t, x, \dot{x}) $.  
Assuming invariance of (18),  we obtain 
\begin{gather}
\eta_{  t t} + ( 2 \eta_ { x t} - \xi_{  t t} ) \dot{x} + ( \eta_{ x  x} - 2 \xi_{  x t} ) {\dot{x}}^2  \\ \nonumber 
-  \xi_{  x x} {\dot{x}}^3 + (( \eta_{ x} - 2 \xi_{  t} ) - 2 \xi_x  \dot{x} ) f( t, x, \dot{x} ) 
- \xi  f_t ( t, x, \dot{x} ) - \eta   f_x  ( t, x, \dot{x} ) + ( \eta_ t + ( \eta_x - \xi_t ) \dot{x} - \xi_x { \dot{x} }^2 ) f_{ \dot{x} } (  ( t, x, \dot{x} ) \equiv 0 
\end{gather} 

 Each $ \xi $ and $ \eta $ here  satisfies above relation for higher dimensional algebra.  Since,    $  f( t, x, \dot{x} ) $  is  polynomial  in  $ \dot{x} $, it yields set of PDE s in $ \xi $ and $ \eta $.
Substituting equation (31-32) into (35)   and separating  null coefficients of  powers of $ \dot{x} $, we obtain;
\begin{gather} 
\xi_{ x x} = -  \alpha_2 \xi_x  \\
\eta_{ x x}  - 2 \xi_{ x t} - 2 \alpha_1 \xi_x = \alpha_2 \eta_x + {\alpha_2}_t \xi + {\alpha_2}_x \eta \\
2 \eta_{ x t} - \xi_{ t t} - 2 \alpha_0 \xi_x - \alpha_1 \xi_t - {\alpha_1}_t \xi - {\alpha_1}_x \eta = 2 \alpha_2 \eta_t \\
\eta_{ t t} - \alpha_0 ( 1 \xi_t - \eta_x) - {\alpha_0 }_t \xi -{\alpha_0}_x \eta_t = 0 
\end{gather} 
The general solution of this homogeneous linear system can be formally written as a superposition of linearly  independent basis solutions $ \xi_a( t,x) $ and $ \eta_a( t,x) $ 
with $ a = 1,2, \cdots r \le 8 $ following Einstein dummy index
\begin{gather} 
 \xi( t, x) = x^a \xi_a ( t,x) \\
 \eta( t, x) = x^a \eta_a( t, x) 
 \end{gather} 
to construct structure constant. Thus symmetry generator takes the form 
\begin{gather} 
g_a( t, x) = \xi_a ( t,x) + \eta_a ( t,x) 
\end{gather}
subject to Cartan killing condition 
\begin{gather} 
[ g_a ( t, x), g_b( t, x ) ] ={ f_{a b} }^c g_c ( t,x ) 
\end{gather} 
where $ { f_{a b} }^c $ is corresponding structure constant .   
 
So, full symmetry group of the differential equation ( 20) should admit  following identities \cite{davis} 
\begin{gather} 
{f_{ ab}}^c \xi_c = [ \xi_a, \xi_{ b t} ] + [ \eta_a , \xi_{ b x} ] \\
{f_{ ab}}^c \eta_c =  [ \xi_a, \eta_{ b t} ] + [ \eta_a , \eta_{ b x} ] \
\end{gather}
Commutation relations are regular  for regular range of values of $ \alpha$ s . So, we assume  initial data to be regular. 
Since x= 0 can not be a regular point, we use 
  initial data at  regular point  is ( t ,x ) =  $ ( t_0, x_0) $ to evaluate structure constants.  This non singular initial data will uniquely determine structure constant. 
In order to represent algebra, we introduce parametrization 
\begin{gather} 
 x^1 = \xi(t_0, x_0) 
 x^2 = \eta(t_0, x_0)  \\
 x^3 = \xi_t( t_0, x_0) 
 x^4 = \eta_x( t_0, x_0) \\
 x^5 = \eta_x ( t_0, x_0) 
 x^6 = \eta_t ( t_0, x_0 ) \\
 x^7 = \cfrac{1} {2} \xi_{ t t} ( t_0, x_0) 
 x^8 =   \cfrac{1} {2} \eta_{ x x } ( t_0, x_0 )  
  \end{gather} 
 According to above relation,  we can adopt 
 \begin{gather}  
 \xi_a ( t_0, x_0) = \delta_{ a 1} ;  \hspace{2 pt} 
  \eta_a( t_0, x_0) = \delta_{ a 2}  \\
  \xi_{ a t} ( t_0, x_0) = \delta_{ a 3} ;  \hspace{2pt} 0 
 \eta_{ a x} ( t_0, x_0) = \delta_{ a 4}  \\
  \xi_{ a x} ( t_0, x_0) = \delta_{ a 5};  \hspace{2pt}                                                                                                                                                                                                                                  
 \;  \eta_{ a t} ( t_0, x_0) = \delta_{ a 6}  \\
  \xi_{ a t t} ( t_0, x_0 ) = \delta_{ a 7}  ; 
 \eta_ { a x x} ( t_0, x_0 ) = 2 \delta_{ a 8} 
 \end{gather} 
 or  
\begin{gather} 
{f_{ ab}}^1  = [\delta_{ a1} ,  \delta_{ b 3} ] + [ \delta_{ a 2}  , \delta_{ b 5} ] \\
{f_{ ab}}^2  =  [ \delta_{a 1} , \delta_{ b 6} ] + [ \delta_{ a 2}  , \delta_{ b 4} ] \\
\end{gather} 
Because of Lie - Cartan integrability conditions, killing equations ( 50 - 60) satisfy Lie algebra.
 Solving above equations simultaneously, we find that equation  (18)  admits sl( 3, $ \mathbb{R} ) $ algebra with following conditions 
 \begin{gather}  
\cfrac{( 1 - q0^2) }{m0} =  - \cfrac{ d r m}{ k p}  \\
\cfrac{s }{ k c^2 } = n 
\end{gather} 
where n presents positive integer. 
Under transitive algebra,  we consider all elements of $ g = \sum_i c_i \cfrac{ \partial} { \partial x_i } $ and  higher order terms 
of $ \mathrm{g} $ for any $ x_i $.

  So, we propose symmetry algebra admitted by non linear  ODE (18) as 
 \begin{prop} 
 We call this symmetry "Hidden Dynamical Symmetry " as this evolves during  dynamical evolution in the network system in terms of linear relation between  immune growth rate and interaction rate i.e 
 \begin{gather} 
 \cfrac{s }{ k c^2 } = n 
 \end{gather} 
\end{prop} 
for any integer value n. 
We  implement most two significant  methods symmetric sub  algebra to integrate ODE and obtain analytical structure of the solution.
The first method  is called method to obtain Normal form of generators in the space of variables or quadrature \cite{cari} .  The second method involves normal form of generators in the space of 
differential invariant function or first integral function \cite{pailas}. This method can be used when number of symmetries is higher or equal to the order of the equation. 
It is necessary to use the generators for the integration procedure in a specific order. This depends on the properties of the algebra. 
When there is a solvable sub algebra of dimension equal to the order of the equation. 
  Integration is performed in correct order and the solution is given solely in terms of quadratures. 

The derived algebra of a Lie algebra $ ( \mathrm{g} , [\cdot, \cdot ] ) $ is the sub algebra $ g^1 $ of $ \mathrm{g} $ , defined by
\begin{gather}
\mathrm{g} ^1 = [ \mathrm{g} , \mathrm{g} ] 
\end{gather} 
while the derived series is the sequence of Lie sub algebra defined by $ {\mathrm{g}}^0 = \mathrm{g} $ and 
\begin{gather} 
\mathrm{g}^{ (k+1)} = [ \mathrm{g}^k, \mathrm{g}^k ] 
\end{gather} 
for any $ k \in \mathbb{N} $ . Such a sequence satisfies $ \mathrm{g}^{(k+1)} \subset \mathrm{g}^k $
and the Lie algebra $ \mathrm{g} $ is said to be solvable if the derived series eventually arrives at the zero sub algebra. 
And  n-  level solvable  algebra  admits  series of invariant  sub algebras defined by
\begin{gather} 
g \equiv g^{(0)}  \supset g^{(1) } \supset g^{(2)}  \cdots  g^{(n-1)}  \supset g^{(n) } \equiv \{0 \} 
\end{gather}                                                                                                                                                                                                                                                                                                                                                                                                                                                                                                                                                                                                                                                                                                                                                                                                                                                                                                                                                                                                                                                                                                                                                                                                                                                                                                                                                                                                                                                                                                                                                                                                                                                                                                                                                                                                                                                                                                                                                                                                                                                                                                                                                                                                                                                                                                                                                                                                                                                                                                                                                                                                                                                                                                                                                                                                                                                                                                                                                                                                                                                                                                                                                                                                                                                                                                                                                                                                                                                                                                                                                                                                                                                                                                                                                                                                                                                                                                                                                                                                                                                                                                                                                                                                                                                                                                                                                                                                                                                                                                                                                                                                                                                                                                                                                                                                                                                                                                                                                                                                                                                                                                                                                                                                                                                                                                                                                                                                                                                                                                                                                                                                                                                                                                                                                                                                                                                                                                                                                                                                                                                                                                                                                                                                                                                                                                                                                                                                                                                                                                                                                                                                                                                                                                                                                                                                                                                                                                                                                                                                                                                                                                                                                                                                                                                                                                                                                                                                                                                                                                                                              
And derived algebra can be  constructed  by some linearly independent sub set of elements of the commutator of the algebra described  by

\begin{gather} 
[ g^{(1)}, g^{(0)} ] \subseteq g^{(1)} 
\end{gather} 
Because of  above relation,  the sub algebra $ g^{(1)} $ is an invariant sub algebra of $ g^{(0)} $. 
Since generators have cardinality l= 8,  they  must follow projective group algebra under   $ sl (3,   \mathcal{R}) $ group.  

 \section{ Normal Form of Generators in the space of Variables  }  
 The first integration method to reduce the order of the ODE into quadrature form or equivalently to find out normal form of generators and use those 
 generators to reduce the order of the equation. 
Under maximal algebra calculation, three generators  follow $ A_{ 3,3} $ solvable sub algebra given by 
  \begin{gather} 
  [ g_5, g_7 ] = 0 \\
[ g_5, g_2] =   g_5 ; 
 [ g_7, g_2] = g_7  
 \end{gather} 
which belongs to  Weyl group, semi direct product of  time dilation and translations $ D \otimes x_2 $.  
Corresponding ub algebra can be identified as 
\begin{gather} 
[ g^{(1)} , g^{(0)} ]  \subseteq g^{(1)} 
\end{gather} 
with 
 \begin{gather} 
 A_{ 3,3} := \lbrace g_5 , g_7; g_2 \rbrace
 \end{gather} 
 Following \cite{patera}, we compose sub algebra as semi direct sums of a one dimensional  sub algebra 
 an abelian ideal with  $ e_1 , e_2, e_3 $ are the bases.  For a Lie algebra $  \mathcal{g} $ with corresponding Lie group $ G = < exp \mathbb{g} > $,  
 the sub algebra can be constructed as semi direct sum of two algebras.
The above (64 -67 ) is an  two level solvable algebra  with $ g^{(1)} = \{ g_5, g_7 \} $ and $ g^{(2)} = \{ 0 \} $ .  
  A   two level solvable algebra is designated by 
  \begin{gather} 
  {0} \equiv g^{(2)} \subset g^{(1)} \subset g^{(0)} 
  \end{gather} 
Following theorem 3 of  \cite{pailas},    first integration method can be repeated n times upon the following chain of cosets,
  \begin{gather} 
g^{(n-1)} \equiv {B_n}^ {(n-1)} \xrightarrow[ ] { pr}  \cdots   \xrightarrow[ ] { pr}  {B_1}^ {0}
\end{gather} 
 where coset is defined  between two derived algebras $ g^{(i)} $ and $ g^{(j)} $  as
 \begin{gather} 
 {B_(j)}^ {(i)} = g^{(i)} - g^{(j)} 
 \end{gather} 
 Here
 \begin{gather} 
 {B_{ (2)}}^{(1)}  = \{ g_5, g_7 \} \\
 [  {B_{ (2)}}^{(1)} , g^{(0)} ] \subseteq { 0} 
 \end{gather} 
 forms abelian algebra. Following theorem 1 \cite{pailas} , reduced generators form an algebra with structure constants a subset of the original ones such that 
 \begin{gather} 
 [  {B_{ (1)}}^{(0)} ,  {B_{ (1)}}^{(0)} ] = {0} 
 \end{gather} 
  The two generators of the coset act transitive. Using this,  transform generators  into normal form. 
   By  introducing new coordinates ( T, X) such that T is the independent variable and X is corresponding dependent variable in normal form. 
as  $  T   =  f( t, x ) $ and $  X = g(t , x ) $ , 
 we employ
\begin{gather} 
g = \{  a  g_5 ;   b   g_7  \} 
\end{gather} 
two dimensional sub algebra 
such that 
\begin{gather} 
\tilde{g}   T     = 0, \tilde{g}   ( X T )     = 1 \\ 
\end{gather} 
holds true. 
 Using twisted Goursat algorithm for decomposable algebra  \cite{patera}  , we obtain a= 1, b=3  which yields 
 \begin{gather} 
 T =   x(t) e^{ - \cfrac{ ( -\alpha_1   + \sqrt{ \alpha_1 ^2 + 4 \alpha_0} )t} {2} }  \\
 X = \cfrac{ x(t) e^{  \cfrac{ ( \alpha-1  + \sqrt{{\alpha_1}^2 + 4 \alpha_0} )t} {2} } } { 4 \sqrt{ {\alpha_1} ^2 + 4 \alpha_0} } 
 \end{gather} 
In terms of normal variable ( canonical)  ,we obtain ODE
\begin{gather} 
 ( {\alpha_1} ^2 + 4 \alpha_0  ) T^3 X_{ T,T} X^2  = 0 
\end{gather}
which is in quadrature form with linear form of solution. 
 Inverse mapping of variables yields solution structure  of target component as
 \begin{gather} 
x_{lin}(t) = \cfrac{ C_2 } { C_1 e^{ \cfrac{ \sqrt{ {\alpha_1}^2  + 4 \alpha_0 } - \alpha_1 )t} {2} }  + \cfrac   { e^{ \cfrac{ \sqrt{ {\alpha_1}^2  + 4 \alpha_0 } - \alpha_1 )t} {2} } } { \sqrt{ {\alpha_1}^2 + 4 \alpha_0 } } } 
\end{gather}
with $ C_1, C_2 $ to be determined from initial conditions.  \\

 \section{ Group Invariant Solution  Structure}
 This method uses differential invariant of the sub group  to obtain solution.    We use abelian solvable sub algebra 
 \begin{gather} 
 [ g_5, g_7 ] = 0 
 \end{gather} 
 under prolonged group to evaluate invariant differential  function in the evolution dynamics.  We use $ g \equiv g_7 $ here. 
 \begin{gather} 
 g^{[1]}  \equiv  \xi(t, x) \cfrac{\partial} {\partial t} + \eta(t,x) \cfrac{\partial}{ \partial x} + \eta^{[1]} \cfrac{ \partial}{ \partial \dot{x}} 
 \end{gather} 
 with 
 \begin{gather} 
 \xi(t, x) = \cfrac{e^{ - ( \cfrac{  \alpha_1  +  \sqrt{ {\alpha-1}^2 + 4 \alpha_0 } )t }{2}}  } { x(t)} \\
 \eta( t, x )  = -  \cfrac{ (  \sqrt{ {\alpha_1}^2 + 4 \alpha_0} - \alpha_1 )}{2} e^{ - ( \cfrac{   \sqrt{ {\alpha_1}^2 + 4 \alpha_0 } - \alpha_1)t }{2}} \\
\eta^{[1]} = \eta_t + ( \eta_x- \xi_t) \dot{x} - \xi_x {\dot{x}}^2  
 \end{gather} 

with  characteristic equation
 \begin{gather} 
 \cfrac{ d t}{ \xi} = \cfrac{ d x} { \eta} = \cfrac{ d \dot{x}} { \eta^{[1]} } 
 \end{gather} 
 which provides two invariant functions , namely $ \phi (t,x) $ and $ \psi( t, x, \dot{x} ) $ with 
 \begin{gather} 
 g \phi(t,x) = 0 ;  \\
 g^{[1]}  \psi( t, x, \dot{x} ) = 0 \\
 g^{[2]} \cfrac{ d \psi}{ d \phi} = 0 
 \end{gather} 
 We concentrate on the system where number of equations in the system is same as number of dependent variables. That means 
 \begin{gather} 
 \cfrac{d}{ d t} = \cfrac{ \partial} {\partial t} + \dot{x}   \cfrac{ \partial} { \partial x } + \ddot{x} \cfrac{\partial } { \dot{x} } 
 \end{gather} 
Two invariant functions gives relation  
 \begin{gather} 
 \cfrac{ d \psi}{ d \phi} = \cfrac{ \psi_t} { \phi_t} = \cfrac{ \psi_t + \dot{x} (t) \psi_x + f \psi_{ \dot{t} } } { \phi_t + \dot{x} \phi_x } 
 \end{gather} 
 Implementing equations (82-91), we obtain 
 \begin{gather} 
 x_{lin} (t)  =  \cfrac{ \sqrt{ {\alpha_1} ^2 + 4 \alpha_0} } { C_2 \sqrt{ {\alpha_1}^2 + 4 \alpha_0 } - C_1 e^{ -t (\sqrt{ {\alpha_1} ^2 + 4 \alpha_0} )} }  e^{   \cfrac{ (-  \alpha_1  + \sqrt{ {\alpha_1} ^2 + 4 \alpha_0 } )t}{2} } 
 \end{gather} 
 where $ C_1 $ and $ C_2$ will be determined from initial conditions.

 \section{Numerical Simulation \& Results } 
  In the model, parameter  u and v play critical role in transition  of the dynamics. So,
we expand solution around linear (analytical) value by adding nonlinear part in a perturbative way i.e 
 
 \begin{gather} 
 x(t) =  x_{lin} (t)  + \epsilon f(x,t)  + \epsilon^2 \cfrac{\partial f (x,t)}{ \partial t} + \cdots  \\
 f(x,t ) =  \cfrac{ p { x (t) }^u}{ m^v + {x(t)}^v}
 \end{gather} 
 Parameter u act as immune stimulation and  v  as correlation parameter in the network. We keep parameter values same as obtained in dynamical simulation. 
Following is the result of simulaiton. Linear behavior is exhibited in the simulation using analytical solution without perturbation ( FIG 6).
Simulation data from linear structure of the solution exhibit presence of  local critical attractor in stable manifold.

 \begin{figure} 

  \centering 
\subfigure {
\epsfig{file=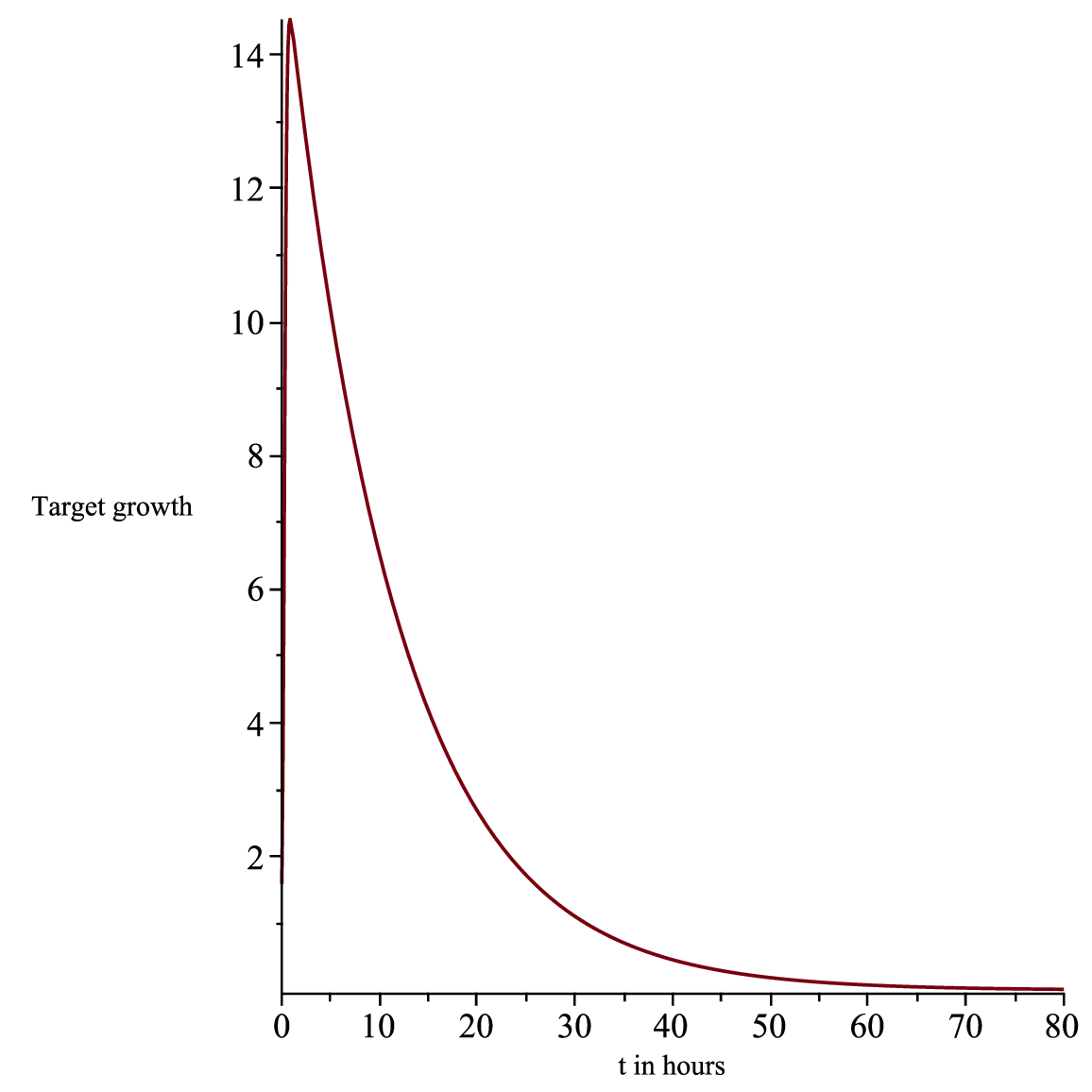, scale=0.25}}
\hspace{8pt} 
\subfigure {
\epsfig{file=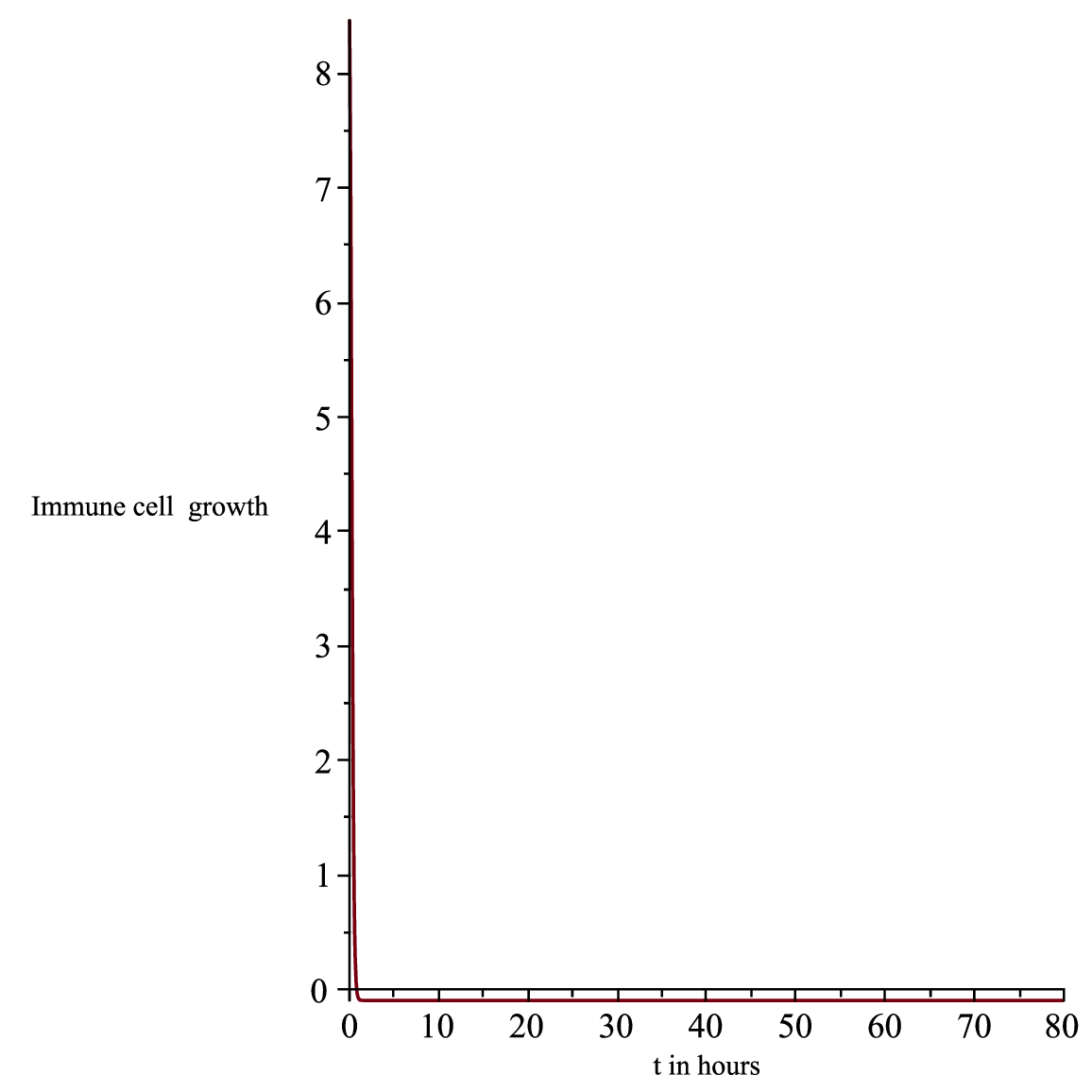, scale=0.25}} 
\hspace{8pt} 
\subfigure {
 \epsfig{file=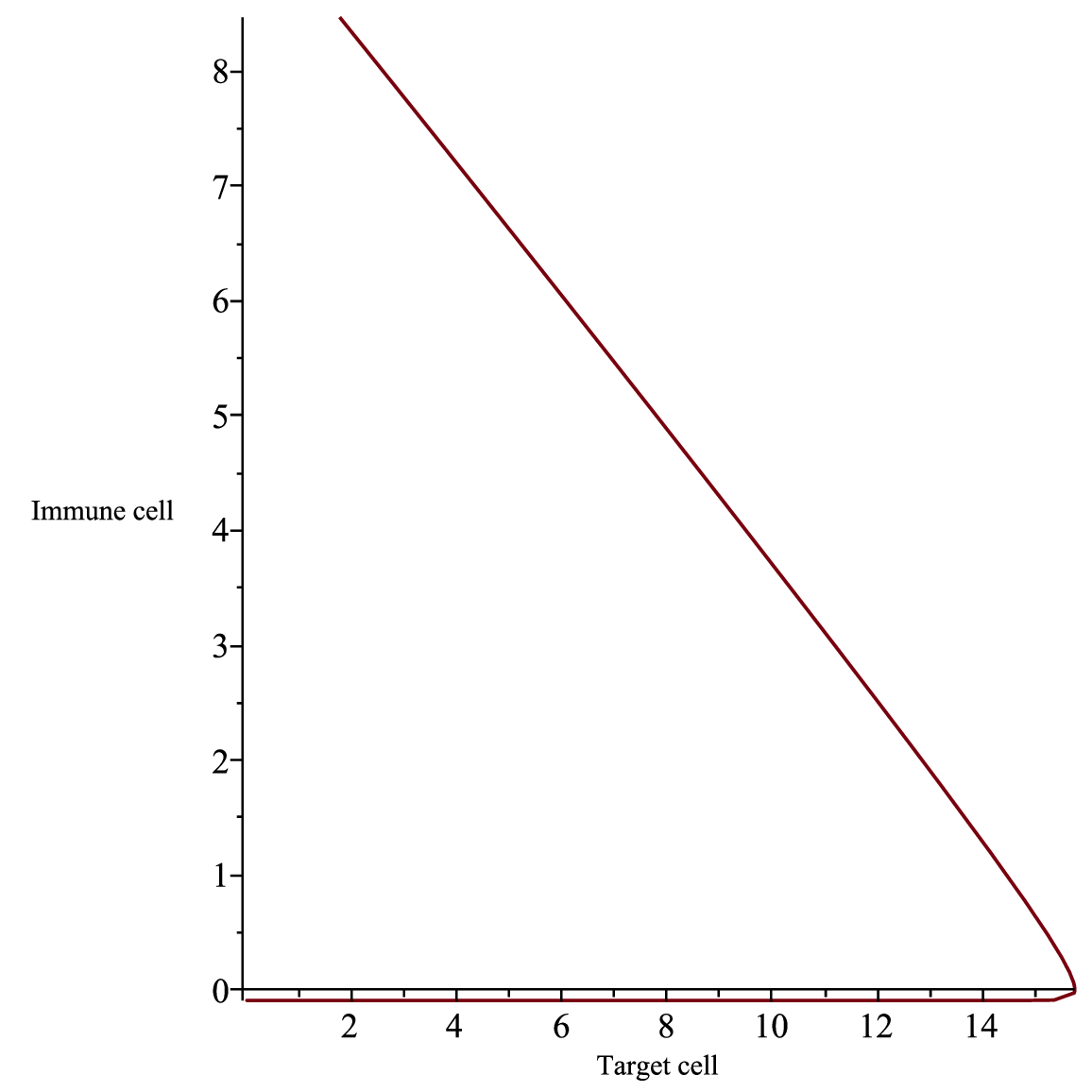, scale=0.25}} 
 \caption{ Linear solution behavior based on analytic structure ; d=1.0, r= 0.701, p=0.642, s= 1.23 , m=- 0.14, c=1.0  k =0.93, u=v=0 } 
 \end{figure} 
 \begin{figure} 
  \centering 
\subfigure {
\epsfig{file=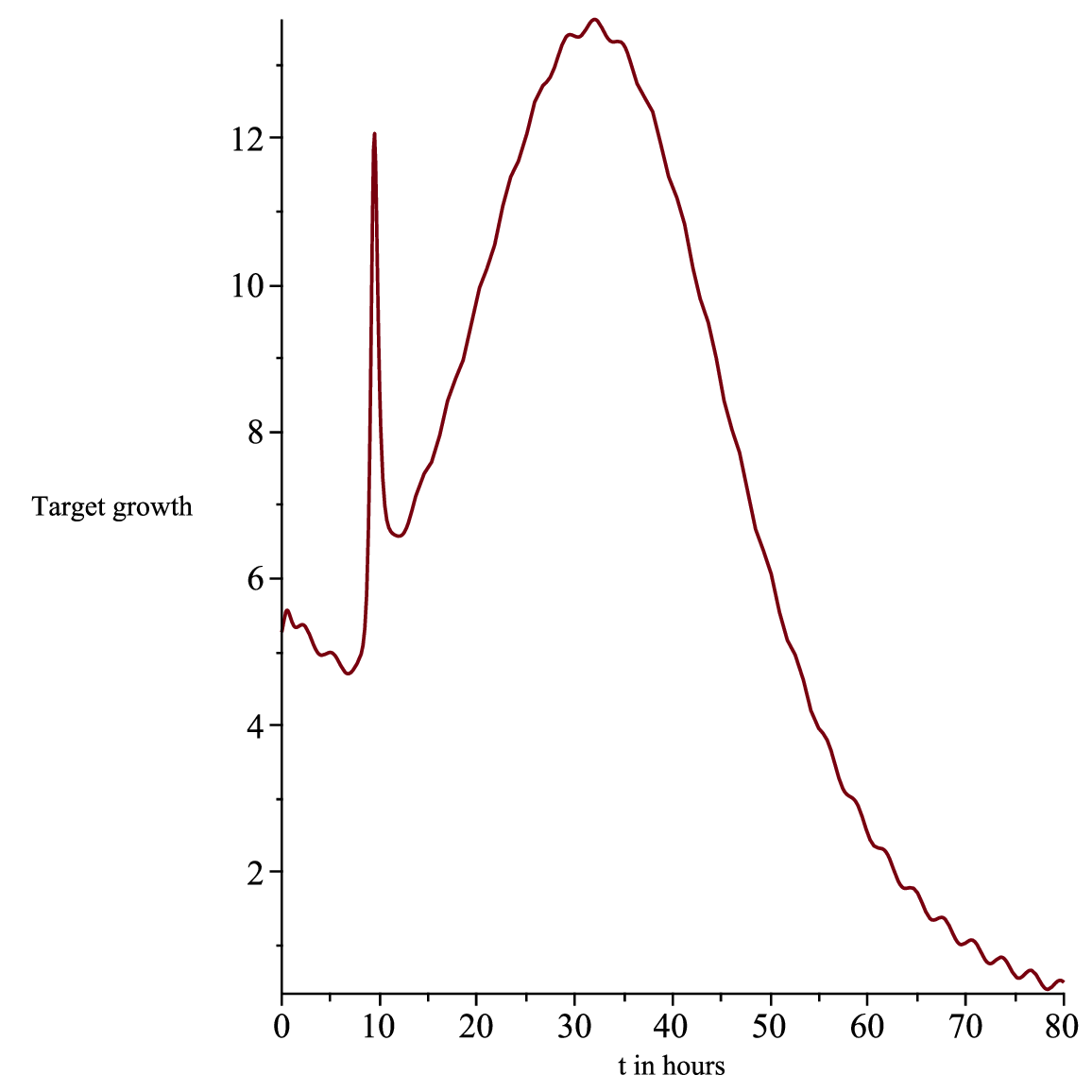, scale=0.25}}
\hspace{8pt} 
\subfigure {
\epsfig{file=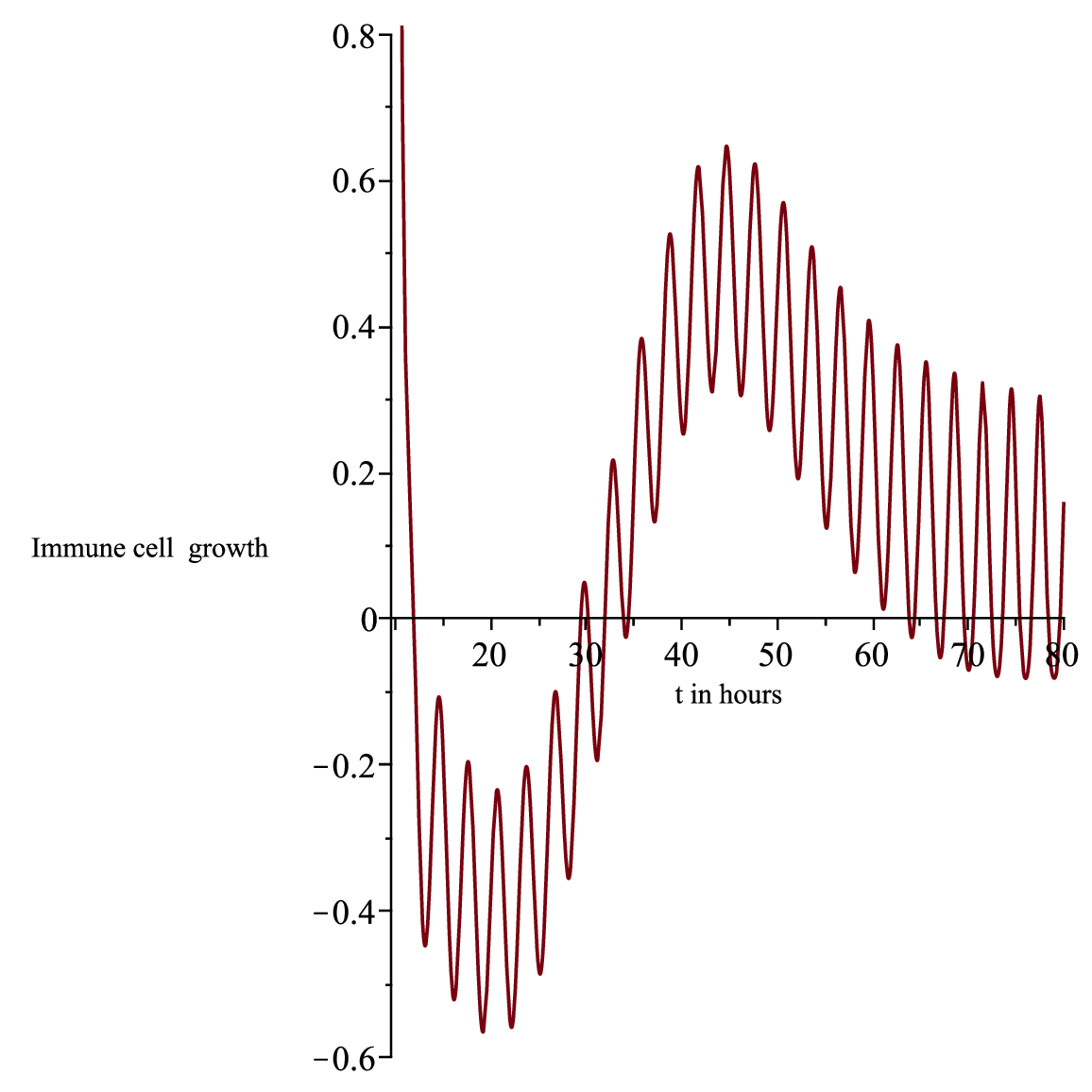, scale=0.25}} 
\hspace{8pt} 
\subfigure {
 \epsfig{file=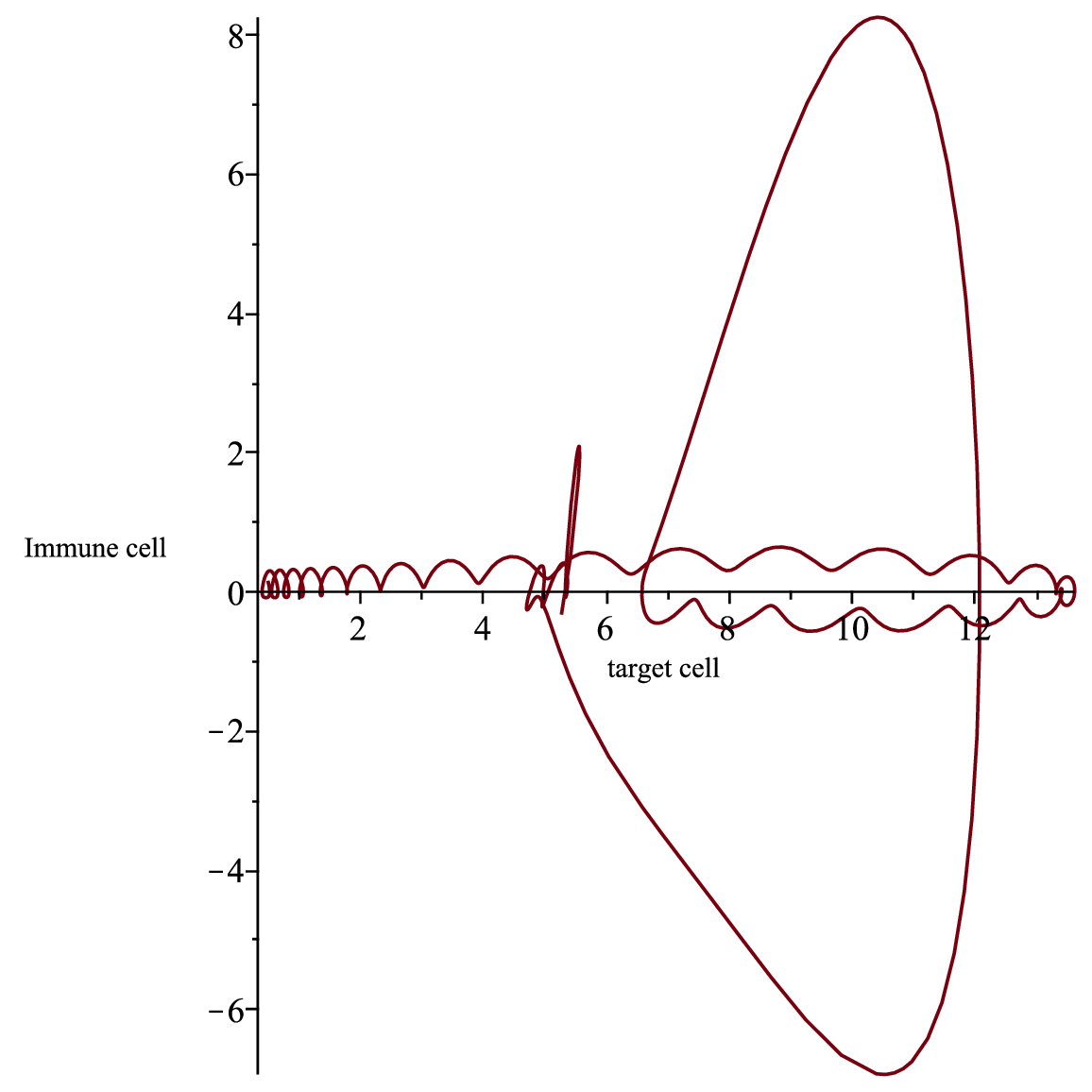, scale=0.25}} 
 \caption{ Dynamical behavior with weak correlation u=v=1; d=1.0, r= 0.701, p=0.642, s= 1.23 , m=-0.142, c=1.0  k =0.93, u=v=1 } 
 \end{figure}  
Setting  u=1, v=1 and second  order perturbation,  growth pattern exhibits presence of limit cycle determined by parameter values 
  d=1.0, r=0.7, p=0.642, c=1.0, k=0.25 ( FIG 7). 
The negative of  (m = -0.142) designates attractive phase of dynamics with stable oscillation in closed manifold.  The target growth shows consistent increase with
numerous small oscillations whereas immune response indicates strong oscillation between 35 and 80 hours. In the simualtion time can be hours or days depending 
on the intensity of infection.  In the first phase, negative oscillatory values of immune density indicates not responding to annihilate target or failure to respond in target invasion.  The way immune responds in any case of target invasion in the body is through antibody protein production via CD+8 or CD+4 cytokinase protein.
The process of production of these protein involes multi step interaction via several meta stages.

Value  u=1, v=3  indicates  strong correlation  between immune and  target dynamics within local domain [FIG 8]. 
Our simulation  reflects chaotic correlated dynamics between two components using second order perturbation.
This shows the role of parameter v as correlation  in non linear growth dynamics. 
\begin{figure}
 \centering 
\subfigure {
\epsfig{file=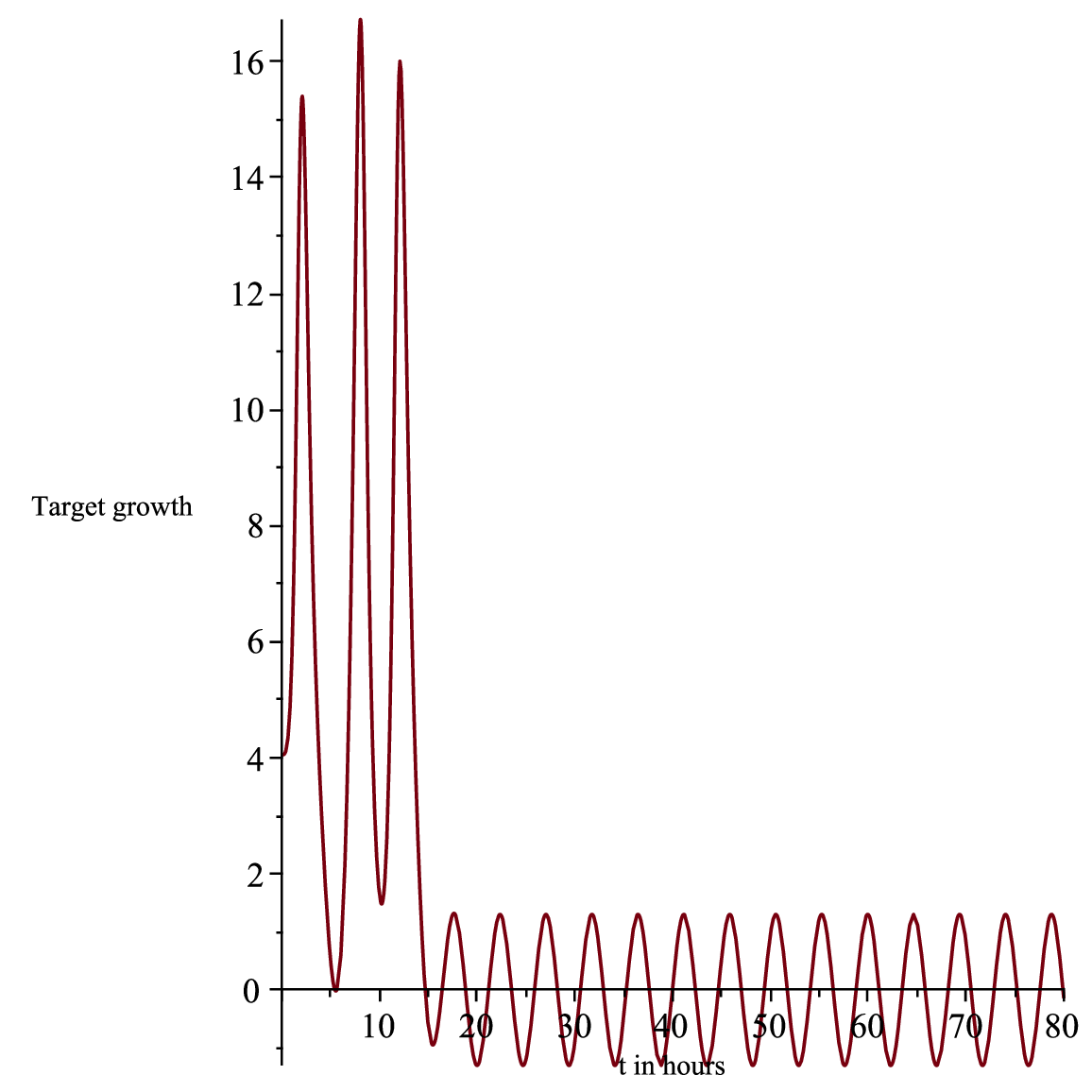, scale=0.25}}
\hspace{8pt} 
\subfigure {
\epsfig{file=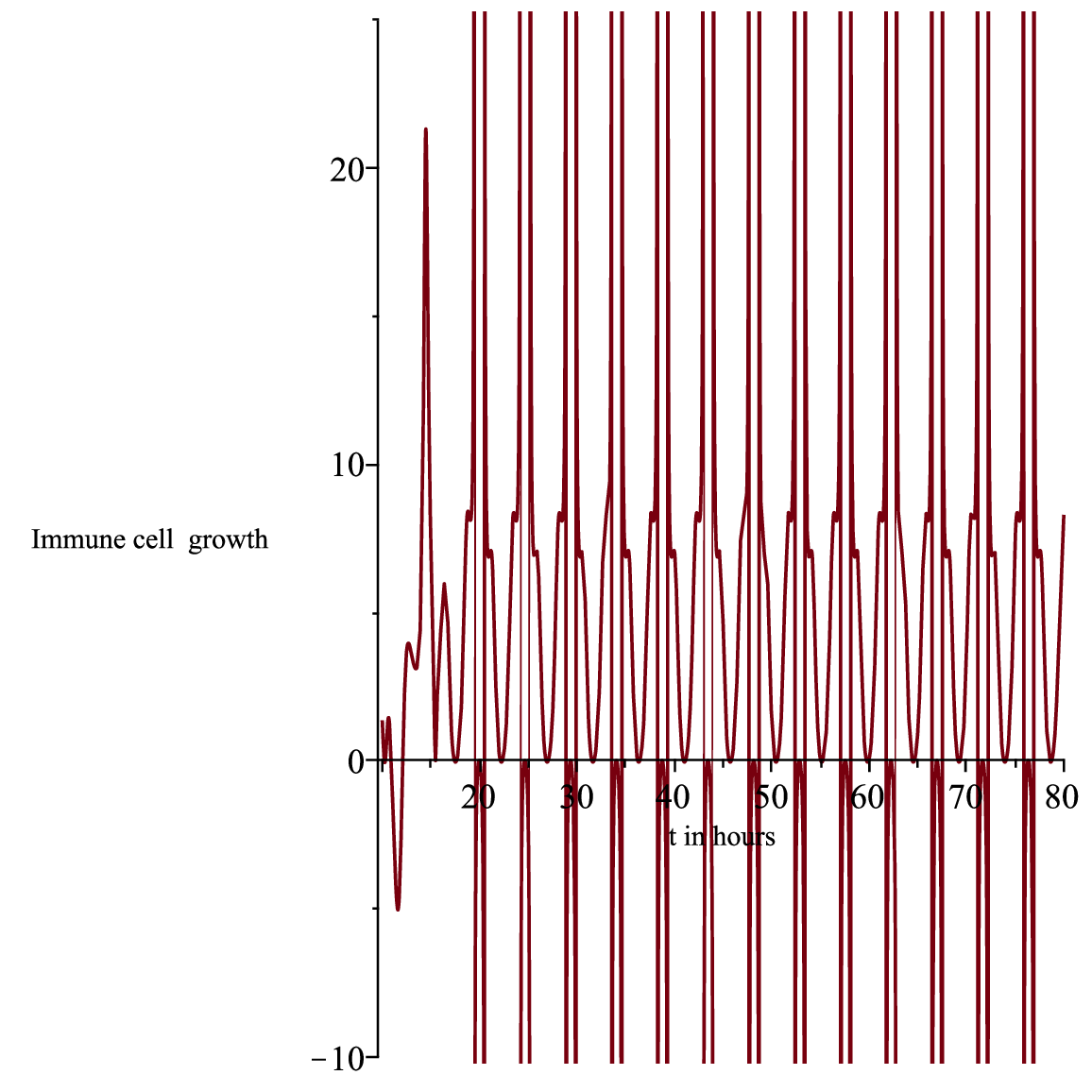, scale=0.25}} 
 \hspace{8pt} 
 \subfigure {
\epsfig{file=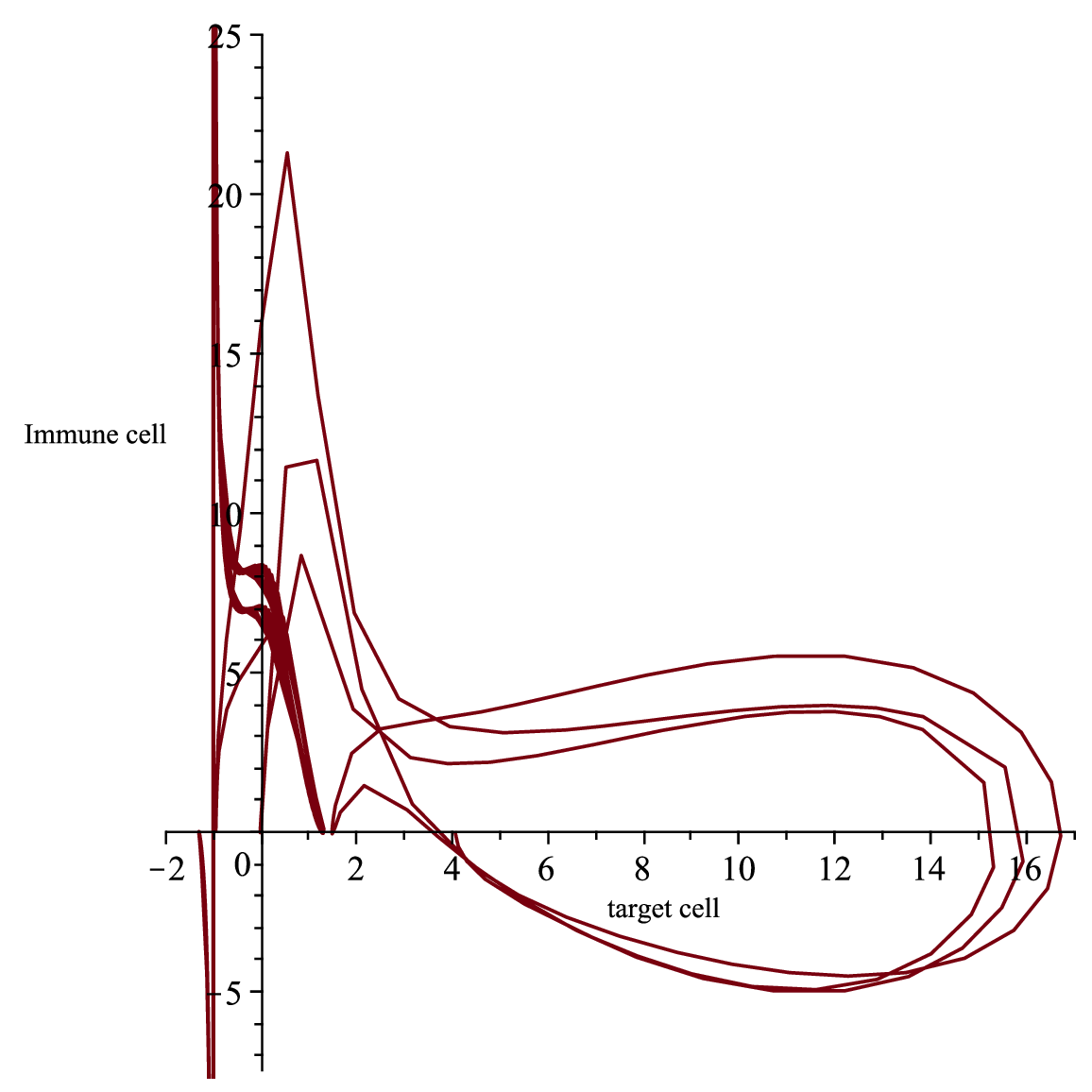, scale= 0.25}}
 \caption{Chaotic behavior of strongly correlated dynamics; d=1.0, r=0.701, p=0.642, s=1.634 , c=1.0,  k=0.15, m= -0.13 }   
 \end{figure}  

Under second order perturbation,  pattern behavior shifts  away from critical attractor designated by positive value of m  ($ m >  0.25 $). 
 As a result, phase trajectory becomes open manifold shown in FIG 9 (therapeutic intervention phase).
Existence of open manifold case [FIG 9] with low immune cell growth indicates that disease persists and requires therapeutic intervention.  
his region is also viable for adaptive immunity in case of lethal infection. The parameter value $ m \ge 0.25 $ plays significant role in development 
 of adaptive immunity which is noted by certain T - cell proliferation through antigen production.  This feature is prominent in both approaches. 
The parameter m here plays  significant role via antigen production in case of adaptive immunity. The existence of continuous oscillation or 
unstable dynamics is noted far away from stable equilibrium ( k= 3.4189 )  when we increase death rate of immune cell d= 1.18 .
When we set  parameter values v=2.0 ,m=0.3, second order perturbation simualtion of  target growth exhibits steady  positive for
 longer time period with maximum value
around 40 hours. Positive value of m drives the system away from closed manifold.   Immune growth pattern within 30 hours of infection can
 not compete with target and then starts target annihilation process 
between 40 and 80 hours after first infection.  
\begin{figure}
 \centering 
\subfigure {
\epsfig{file=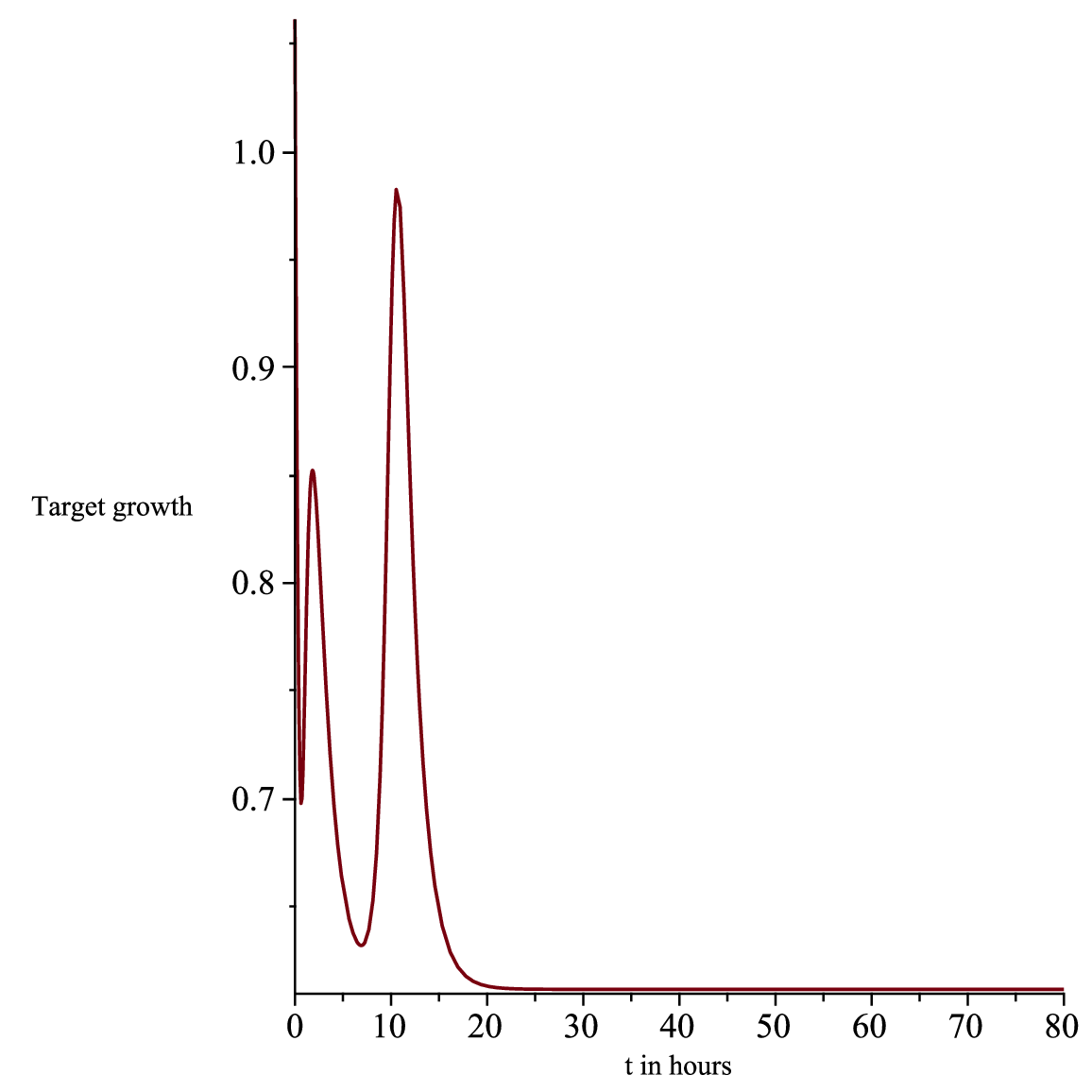, scale=0.25}}
\hspace{8pt} 
\subfigure {
\epsfig{file=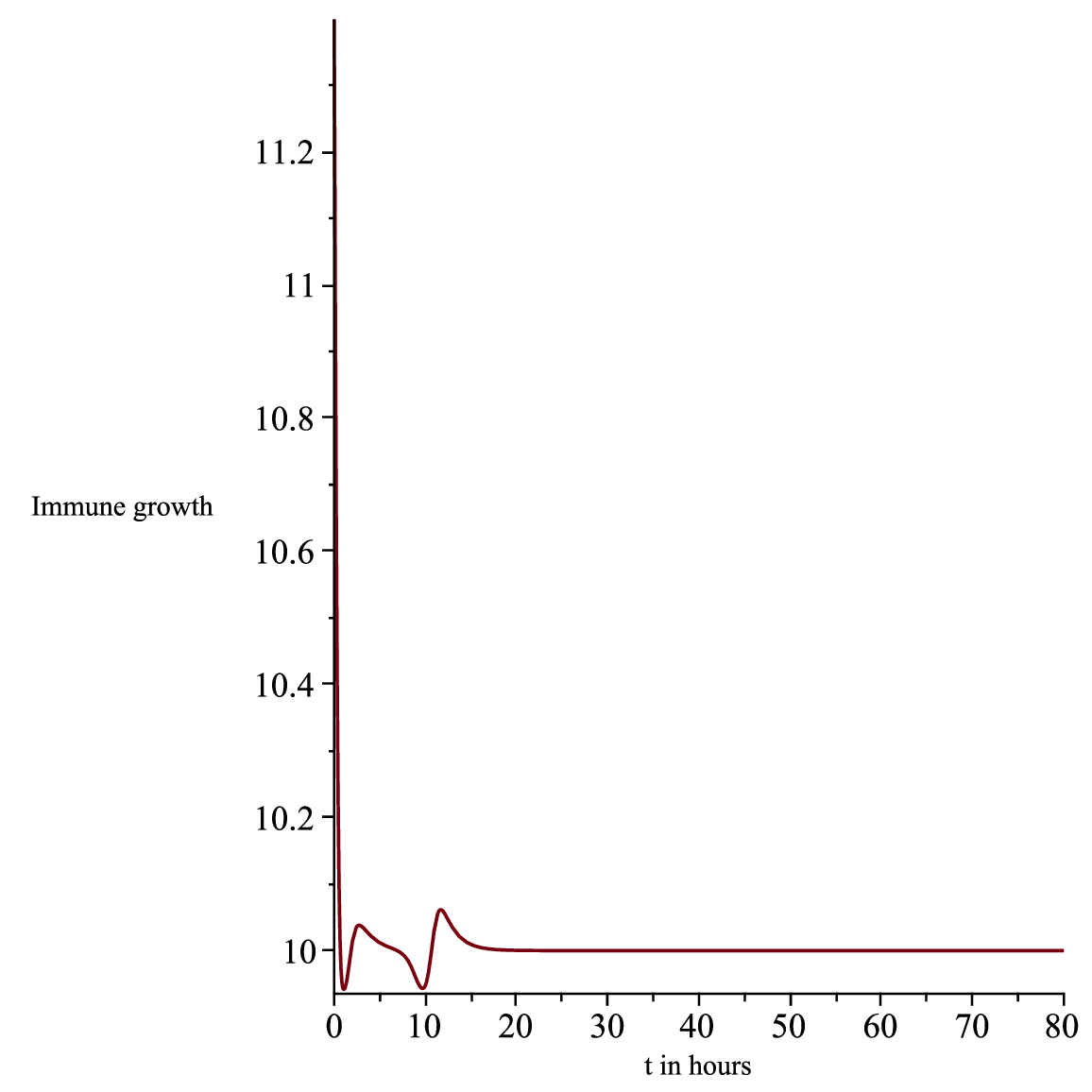, scale=0.25}} 
\hspace{8pt} 
\subfigure {
 \epsfig{file=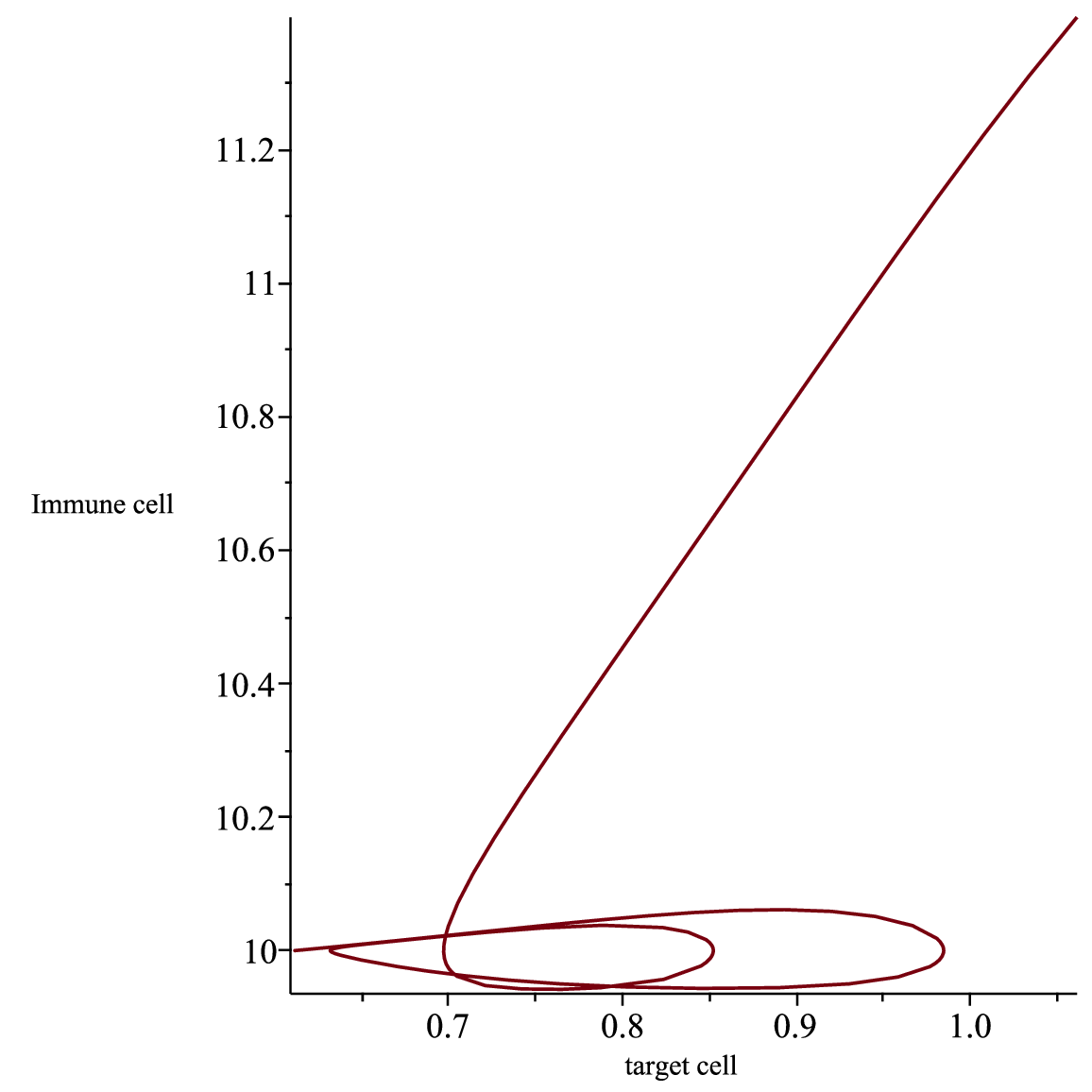, scale=0.25}} 
 \caption{Pattern behavior far away from critical attractor ;  $ m \ge 0.25 $, d=1., r= 0.701, p=0.652, s= 1.625 ,c=1.0,  u=1.0 , v=3.0, k= 0.254} 
 \end{figure}


\newpage 
\section{Appendix} 

\begin{table}[ht] 
\caption{ Symmetry Generator following sl( 3, $ \mathbb{R} $ )  sub algebra  of the ODE } 
\centering 
 \begin{tabular}{|c|c|}  \\
 \hline 
$  g_1 $  & $  \partial_t  $ \\ \hline \hline 
 $  g_2  $ & $  x(t)  \partial_x $ \\ \hline \hline
$  g_3 $ &  - $ x(t)^2  e^{ ( \cfrac{  \alpha_1 + \sqrt{ {\alpha_1}^2 + 4 \alpha_0 } )t }{2}}   \partial_x $  \\ \hline 
$ g_4 $ &   $  x(t)^2   e^{ ( - \cfrac{  \alpha_1  - \sqrt{ {\alpha_1}^2 + 4 \alpha_0 } )t }{2}}  \partial_x $ \\ \hline \hline
$ g_5 $ &  $ \cfrac{e^{  ( \cfrac{  \alpha_1  - \sqrt{ {\alpha_1}^2 + 4 \alpha_0 } )t }{2}}  } { x(t)} \partial_t  + \cfrac{ ( \alpha_1  + \sqrt{ {\alpha_1}^2 + 4 \alpha_0} )}{2} e^{ - ( \cfrac{   + \sqrt{ {\alpha_1}^2 + 4 \alpha_0 } - \alpha_1)t }{2}} \partial_x $ \\ \hline 
$ g_7 $ & $  \cfrac{e^{ - ( \cfrac{  \alpha_1  +  \sqrt{ {\alpha-1}^2 + 4 \alpha_0 } )t }{2}}  } { x(t)} \partial_t -  \cfrac{ (  \sqrt{ {\alpha_1}^2 + 4 \alpha_0} - \alpha_1 )}{2} e^{ - ( \cfrac{   \sqrt{ {\alpha_1}^2 + 4 \alpha_0 } - \alpha_1)t }{2}}  \partial_x $ \\ \hline 
\end{tabular} 
\end{table} 

  \begin{table}[ht] 
  \caption{ Model parameter values for primary/secondary /therapeutic Intervention   } 
  \centering 
  \begin{tabular} { | c | c | c |  c | c |c | c |c |c | } 
  \hline \hline
  n & u & v & r & k & p & s  & m & Phase \\ \hline 
  1 & 1 &1 & 0.7 & 0.12 & 0.642 & 1.23 & 0.2 & Linear \\ \hline 
  3 & 1 & 2 & 0.701  & 0.12  & 0.642  &  1.56 &  -0.13 & Periodic   \\ \hline 
  3 & 1 & 3 & 0.702 & 0.265 & 0.642 & 1.65  &   0.26 & Far away from critical attractor  \\ \hline
  \end{tabular} 
  \end{table}

  \section{Acknowlegment} 
  R. Dutta is thankful to Department of Mathematics  for computation support. \\

\section{Financial Support}
 This research received no specific grant from any funding agency, commercial or nonprofit sectors.
 
 \section{Conflict of Interests Statement}
 The authors have no conflicts of interest to disclose.
 
 \section{Ethical Statement}
 This research did not required ethical approval.

  \end{document}